\documentclass[longauth]{aa} 

\usepackage{graphicx}
\usepackage{txfonts}
\usepackage[sc]{mathpazo} 
\usepackage[T1]{fontenc} 
\usepackage{microtype} 

\usepackage{hyperref} 

\usepackage{booktabs} 
\usepackage{float} 

\usepackage{lettrine} 
\usepackage{paralist} 

\usepackage[nomessages]{fp} 
\usepackage{xspace}
\usepackage{xcolor}
\usepackage{longtable}
\usepackage{multirow}
\usepackage{tablefootnote}
\usepackage{tablefootnote}
\usepackage{numprint}
\npthousandsep{\,}
\usepackage{tikz} 

\def \eg {e.g.,\xspace}
\def \ie {i.e., }

\usepackage{natbib}
\bibpunct{(}{)}{;}{a}{}{,} 

\def \dnn {DNN\xspace}  
\def \dnns {DNNs\xspace}

\def \ps {SP\xspace}    
\def \pc {CP\xspace}    

\def \teach {teaching set\xspace}       

\def \sensitivity {True ET Rate\xspace} 
\def \specificity {True LT Rate\xspace} 

\def \zlim {2\xspace}                                   
\def \urplane {2.22\xspace}                             
\def \ur {\ensuremath{u-r}\xspace}              

\def \zurich {ZSMC\xspace}      
\def \photoz {CPhR\xspace}      

\def \sp {S\'{e}rsic index and photometry\xspace}

\def \net {99}                          
\def \nlt {1735}                        
\FPeval{\soln}{clip(\net+\nlt)}
\def \ngal {\soln\xspace}       
\def \nur {1787}
\def \nurgal {\nur\xspace}                      

\FPeval{\soltrain}{round(\soln * 0.7 - 0.5, 0)}
\FPeval{\solval}{round(\soltrain * 0.2, 0)}
\FPeval{\soltest}{clip(\soln - \soltrain)}

\FPeval{\pet}{clip(\net / \soln * 100)} 
\FPeval{\plt}{clip(\nlt / \soln * 100)} 

\def \netC {5737}                               
\def \nltC {28951}                      
\FPeval{\solnC}{clip(\netC+\nltC)}
\def \ngalC {\numprint{\solnC}\xspace}  

\def \soltrainC {5000}
\def \ntrainC {\soltrainC\xspace}       
\FPeval{\solvalC}{round(\soltrainC * 0.2, 0)}
\FPeval{\soltestC}{clip(\solnC - \soltrainC)}
\def \ntestC {\numprint{\soltestC}\xspace}      

\FPeval{\petC}{clip(\netC / \solnC * 100)} 
\FPeval{\pltC}{clip(\nltC / \solnC * 100)} 

\def \cp {Concentration and photometry\xspace}

\def \netc {114}                                
\def \nltc {2178}                       
\FPeval{\solnc}{clip(\netc+\nltc)}
\def \ngalc {\solnc\xspace}     
\def \nurgalc {2189\xspace}             

\FPeval{\soltrainc}{round(\solnc * 0.7, 0)}
\def \ntrainc {\soltrainc\xspace}       
\FPeval{\solvalc}{round(\soltrainc * 0.2, 0)}
\FPeval{\soltestc}{clip(\solnc - \soltrainc)}
\def \ntestc {\numprint{\soltestc}\xspace}      

\FPeval{\petc}{clip(\netc / \solnc * 100)} 
\FPeval{\pltc}{clip(\nltc / \solnc * 100)} 

\def \netCc {9977}                              
\def \nltCc {95781}                     
\FPeval{\solnCc}{clip(\netCc+\nltCc)}
\def \ngalCc {\numprint{\solnCc}\xspace}        

\def \soltrainCc {10000}
\def \ntrainCc {\numprint{\soltrainCc}\xspace}  
\FPeval{\solvalCc}{round(\soltrainCc * 0.2, 0)}
\FPeval{\soltestCc}{clip(\solnCc - \soltrainCc)}
\def \ntestCc {\numprint{\soltestCc}\xspace}    

\FPeval{\petCc}{clip(\netCc / \solnCc * 100)} 
\FPeval{\pltCc}{clip(\nltCc / \solnCc * 100)} 

\usepackage[normalem]{ulem}

\begin{document}

   \title{Galaxy classification: deep learning \\ on the OTELO and COSMOS databases}


   \author{
          \mbox{Jos\'{e} A. de Diego 
          \inst{\ref{inst:unam}, \ref{inst:iac}}}
          \and
          \mbox{Jakub Nadolny 
          \inst{\ref{inst:iac}, \ref{inst:ull}}}
        \and
          \mbox{\'{a}ngel Bongiovanni 
          \inst{\ref{inst:iram}, \ref{inst:aspid}}}
        \and
          \mbox{Jordi Cepa 
          \inst{\ref{inst:iac}, \ref{inst:aspid},\ref{inst:ull}}}
        \and
          \mbox{Mirjana Povi\'{c} 
          \inst{\, \ref{inst:essti}, \ref{inst:iaa}}}
        \and
          \mbox{Ana Mar\'{\i}a P\'{e}rez Garc\'{\i}a 
          \inst{\ref{inst:aspid}, \ref{inst:inta}}}
        \and
          \mbox{Carmen P. Padilla Torres 
          \inst{\ref{inst:iac}, \ref{inst:ull}, \ref{inst:inaf}}}
        \and
          \mbox{Maritza A. Lara-L\'{o}pez 
          \inst{\ref{inst:dark}}}
        \and
          \mbox{Miguel Cervi\~{n}o 
          \inst{\ref{inst:inta}}}
        \and
          \mbox{Ricardo P\'{e}rez Mart\'{\i}nez 
          \inst{\ref{inst:aspid}, \ref{inst:esa}}}
        \and
          \mbox{Emilio J. Alfaro 
          \inst{\ref{inst:iaa}}}
        \and
          \mbox{H\'{e}ctor O. Casta\~{n}eda 
          \inst{\ref{inst:ipn}}}
        \and
          \mbox{Miriam Fern\'{a}ndez-Lorenzo 
          \inst{\ref{inst:iaa}}}
        \and
          \mbox{Jes\'{u}s Gallego 
          \inst{\ref{inst:ucm}}}
        \and
          \mbox{J. Jes\'{u}s Gonz\'{a}lez \inst{\ref{inst:unam}}}
        \and
          \mbox{J. Ignacio Gonz\'{a}lez-Serrano 
          \inst{\ref{inst:ifca}, \ref{inst:aspid}}}
        \and
          \mbox{Irene Pintos-Castro 
          \inst{\ref{inst:toronto}}}
        \and
          \mbox{Miguel S\'{a}nchez-Portal 
          \inst{\ref{inst:iram}, \ref{inst:aspid}}}
        \and
          \mbox{Bernab\'{e} Cedr\'{e}s
          \inst{\ref{inst:iac}, \ref{inst:ull}}}
        \and
          \mbox{Mauro Gonz\'{a}lez-Otero
          \inst{\ref{inst:iac}, \ref{inst:ull}}}
        \and
          \mbox{D. Heath Jones 
          \inst{\ref{inst:newcastle}}}
        \and
          \mbox{Joss Bland-Hawthorn 
          \inst{\ref{inst:sydney}}}
          }

   \institute{
   Instituto de Astronom\'{\i}a,
        Universidad Nacional Aut\'{o}noma de M\'{e}xico,
    Apdo. Postal 70-264, 04510 Ciudad de M\'{e}xico, Mexico\\
    \email{jdo@astro.unam.mx} \label{inst:unam}
                  \and
   Instituto de Astrof\'{\i}sica de Canarias (IAC),
    E-38200 La Laguna, Tenerife, Spain \label{inst:iac}
              \and
   Departamento de Astrof\'{\i}sica,
    Universidad de La Laguna (ULL), 
    E-38205 La Laguna, Tenerife, Spain \label{inst:ull}
              \and
   Instituto de Radioastronom\'{\i}a Milim\'{e}trica (IRAM),
    Av. Divina Pastora 7, Local 20, 
    18012 Granada, Spain \label{inst:iram}
              \and
   Asociaci\'{o}n Astrof\'{\i}sica para la Promoci\'{o}n de la Investigaci\'{o}n,
    Instrumentaci\'{o}n y su Desarrollo, ASPID, 
    E-38205 La Laguna, Tenerife, Spain \label{inst:aspid}
              \and
   Ethiopian Space Science and Technology Institute  
    (ESSTI),
    Entoto Observatory and Research Center (EORC), 
    Astronomy and Astrophysics Research Division, 
    PO Box 33679, Addis Ababa, Ethiopia \label{inst:essti}
              \and
    Instituto de Astrof\'{\i}sica de Andaluc\'{\i}a, CSIC, 
    Glorieta de la Astronom\'{\i}a s/n, E-18080,
    Granada, Spain \label{inst:iaa}
              \and
   Depto. Astrof\'{\i}sica, 
        Centro de Astrobiolog\'{\i}a (INTA-CSIC),
        ESAC Campus, Camino Bajo del Castillo s/n, 
        28692, Villanueva de la Ca\~{n}ada, Spain \label{inst:inta}
              \and
   Fundaci\'{o}n Galileo Galilei, 
        Telescopio Nazionale Galileo, 
        Rambla Jos\'{e} Ana Fern\'{a}ndez P\'{e}rez, 7, 
        38712 Bre\~{n}a Baja, Santa Cruz de la Palma, Spain \label{inst:inaf}
             \and
   DARK, Niels Bohr Institute, 
    University of Copenhagen, 
    Lyngbyvej 2, 
    Copenhagen DK-2100, Denmark \label{inst:dark}
              \and
   ISDEFE for European Space Astronomy Centre (ESAC)/ESA,
    P.O. Box 78, 
    E-28690, Villanueva de la Ca\~{n}ada, 
    Madrid, Spain \label{inst:esa}
              \and
   Departamento de F\'{\i}sica,
    Escuela Superior de F\'{\i}sica y Matem\'{a}ticas, 
    Instituto Polit\'{e}cnico Nacional, 
    M\'{e}xico D.F., Mexico \label{inst:ipn}
              \and
        Departamento de F\'{\i}sica de la Tierra y Astrof\'{\i}sica, 
         Facultad CC F\'{\i}sicas,
         Instituto de F\'{\i}sica de Part\'{\i}culas y del Cosmos, 
         IPARCOS,
         28040 Universidad Complutense de Madrid, 
         Spain\label{inst:ucm}
              \and
   Instituto de F\'{\i}sica de Cantabria
    (CSIC-Universidad de Cantabria), 
    E-39005 Santander, Spain \label{inst:ifca}
              \and
   Department of Astronomy \& Astrophysics,
    University of Toronto, Canada \label{inst:toronto}
              \and
   English Language and Foundation Studies Centre, 
    University of Newcastle, 
    Callaghan NSW 2308, Australia \label{inst:newcastle}
              \and
   Sydney Institute of Astronomy, 
    School of Physics, 
    University of Sydney, 
    NSW 2006, Australia \label{inst:sydney}
%
             }

        \date{Version: \today}

 
  \abstract
   {The accurate classification of hundreds of thousands of galaxies  observed in modern deep surveys is imperative if we want to understand the universe and its evolution.}
   { 
   Here, we report the use of machine learning techniques to classify early- and late-type galaxies in the OTELO and COSMOS databases using optical and infrared photometry and available shape parameters: either the S\'{e}rsic index or the concentration index.}
        {We used three classification methods for the OTELO database: 1) \ur color separation, 2) linear discriminant analysis using \ur and a shape parameter classification, and 3) a deep neural network using the $r$ magnitude, several colors, and a shape parameter.
        We analyzed the performance of each method by sample bootstrapping and tested the performance of our neural network architecture using COSMOS data.}
   {
   The accuracy achieved by the deep neural network is greater than that of the other classification methods, and it can also  operate with missing data.
   Our neural network architecture is able to classify both OTELO and COSMOS datasets regardless of small differences in the photometric bands used in each catalog.}
   {In this study we show that the use of deep neural networks is a robust method to mine the cataloged data.}


\maketitle

%

\section{Introduction}

Galaxy morphological classification plays a fundamental role in descriptions of the galaxy population in the universe, and in our understanding of galaxy formation and evolution.
Galaxy morphology is related to key physical, evolutionary, and environmental properties, such as system dynamics \citep{djo87, ger01, deb06, fal19, rom12}, the stellar formation history \citep{ken98, bru03, kau03, lov19}, gas and dust content \citep[\eg][]{lia19}, galaxy age \citep{ber10}, and interaction and merging events \citep[\eg][]{rom12}.
Early galaxy classifications strategies were based on the visual aspect of the objects, differentiating among spiral, elliptical, lenticular, and irregular galaxy types according to their resolved morphology.
Examples of these strategies are the original classification schemes by \citet{hub26} and \citet{vau59}.
This methodology has reached historic marks during the last decade through the citizen science initiative known as {\it Galaxy Zoo}. 
It stands out for being the largest effort made to visually classify more than \numprint{900000} galaxies from the Sloan Digital Sky Survey \citep[SDSS;][]{fuk96} galaxies brighter than $r_{\rm SDSS} = 17.7$ with proven reliability \citep{lin11}. 
After this milestone, this crowd-sourced astronomy project also included the analysis of datasets from the Kilo-Degree Survey (KiDS) imaging data in the Galaxy and Mass Assembly (GAMA) fields, classifying typical edge-on galaxies at $z<0.15$ \citep{hol19}, and the quantitative visual classification of approximately \numprint{48000} galaxies up to $z\sim3$ in three Hubble Space Telescope (HST) fields of the Cosmic Assembly Near-infrared Deep Extragalactic Legacy Survey \citep[CANDELS;][]{sim17}. 
Using the visual classification approach, the morphology and size of luminous, massive galaxies at $0.3 < z < 0.7$ targeted by the Baryon Oscillation Spectroscopic Survey \citep[BOSS;][]{daw13} of SDSS-III were also determined \citep{mas11} using HST and Cosmic Evolution Survey\footnote{\url{http://cosmos.astro.caltech.edu}} \citep[COSMOS;][]{sco07} data. 

%
However, the availability of larger telescopes and sophisticated instruments has made visual classification unfeasible because most galaxies are barely resolved, making identification of their morphological type very difficult, and the number of discovered galaxies has increased dramatically since the introduction of digital surveys dedicated to probing larger and deeper volumes in the universe.
This issue will be even more critical in the near future when the next generation of large surveys such as the \emph{Large Synoptic Survey Telescope} \citep{tys02} or the results from {\it Euclid} mission \citep{lau11} produce petabytes of information and trigger the need for time-domain astronomy \citep{hlo19} far exceeding the capacity of available human resources to manage this information.
For this reason, the automated classification of galaxies has become an intense area of research in modern astronomy.

Previous research into automated galaxy-classification algorithms  has focused on colors, shape, and morphological parameters related to galaxy light distribution, such as concentration and asymmetry \citep[\eg][]{abr94, ber00, con03, con06, pov09, pov13, pov15, den13}. 
Joint automated and visual classification procedures have been implemented in extragalactic surveys such as for example COSMOS \citep{cas07, zam07} and GAMA \citep{alp15}. 
Another approach involves the fitting of spectral energy distributions (SEDs)  using galaxy templates \citep{ilb09}.
 In a complementary fashion, \citet{str01} investigated the dichotomous classification in early- and late-type (ET and LT) galaxies. 
For these authors, the ET group includes the E, S0, and Sa morphological types, while the LT group comprises Sb, Sc, and Irr galaxies.
Furthermore, using the well-known tendency of the LT to be bluer than the ET galaxies, \citeauthor{str01} propose the \ur color to separate between these galaxy types.
S\'{e}rsic and concentration indexes have also been used, alone or in combination with the \ur color, to separate ET from LT galaxies \citep[\eg][]{con03, kel12, den13, vik15}.
A far more complicated and expensive classification, in terms of computational and observational resources, consists in fitting a set of either empirical or modeled SED templates to the galaxy continuum \citep[\eg][]{col80, kin96}. 
Currently, there are some public codes that are able to perform such template-based classifications \citep[\eg LePhare:][]{arn99, ilb06}.

Classification can be addressed in machine learning through supervised learning techniques, which consist in training a function that maps inputs to outputs learning from input--output pairs, and using this function to assign new observations in two or more predefined categories. 
Supervised learning techniques include decision trees \citep{bar20}, random forests \citep{mil17}, linear discriminant analysis (LDA) \citep{mur87}, support vector machines \citep{hue08}, Bayesian classifiers \citep{hen11}, and neural networks \citep{bal04}, among others.

Machine Learning algorithms are increasingly used for classification in large astronomical databases \citep[\eg][]{abo18}.
In particular, LDA is a common classifying method used in statistics, pattern recognition, and machine learning.
Linear discriminant analysis classifiers attempt to find linear boundaries that best separate the data.
Recently, LDA has being used for galaxy classification in spiral and elliptical morphological types \citep{fer15}, classification of Hickson’s compact groups of galaxies \citep{abd19}, and galaxy merger identification \citep{nev19}.

In recent years, neural networks have become very popular in different research areas because of their ability to perform outstanding accurate classifications, and regression and series analyses.
A typical neural network is made up of a number of hidden layers, each with a certain quantity of neurons that perform tensor operations.
There are several network types which are oriented to solve different issues \citep[a brief explanation of different networks can be found in][]{bar19}.
Also, \citet{bus18} used a one-dimensional convolutional neural network (CNN) for classification and redshift estimates of quasar spectra extracted from the BOSS.
Much of the recent research has focused on two-dimensional CNN  classification of galaxy images \citep[\eg][]{ser96, hue15, die15, dom18, per18, wal20}.

In the future, neural networks will probably gain more importance and become the primary technique for classification of astronomical images.
However, there are two drawbacks that limit the use of CNN in astronomical research at  present.
The first is the network bandwidth, which prevents the download of large amounts of heavy images obtained in remote observatories.
The second drawback is the computational and hardware resources needed to train a two-dimension CNN with tens of thousands of images.

Dense (or fully connected) neural networks (\dnn) are used to solve general classification problems applied to tabulated data. 
In astronomy, \dnns have been applied to morphological type classification in low-redshift galaxies.
Thus, \citet{sto92} designed a simple \dnn architecture for morphological classification of 5217 galaxies drawn from the ESO-LV catalog \citep{lau89} using 13 parameters (most of them geometrical) in five different classes, obtaining an accuracy of 56\%.
\citet{nai95} used the same architecture for 830 bright galaxies ($B \leq 17$) and 24 parameters, reducing the parameter space dimension through principal components analysis.
\citet{ser93} used \dnn autoencoders for unsupervised classification of galaxies into three major classes: Sa+Sb, Sc+Sd, and SO+E. 
\citet{sre18} applied a \dnn to a sample of 7528 galaxies at redshifts $z < 0.06$ extracted from the Galaxy And Mass Assembly survey (GAMA\footnote{\url{http://www.gama-survey.org}}) achieving an accuracy of 89.8\% for spheroid- versus disk-dominated classification.

These earlier works showed that \dnns are capable of performing accurate classification tasks on processed data such as photometry, colors, and shape parameters of low-redshift galaxies \citep{sto92,nai95,bal04}. 
However, compared with image-oriented CNNs, little attention has been paid recently to the use of \dnns for galaxy classification, even if these networks do not require the large quantity of resources used by the CNN.
Moreover, both neural network software development \citep[\eg Tensorflow,][]{aba16} and hardware computation power (both central and graphics processing units) have increased dramatically, boosting the capabilities of \dnn applications.

In this paper we extend the use of \dnns to the morphological classification of galaxies up to redshifts $z \leq 2$. 
We compare the performance of different galaxy classification techniques applied to a sample of galaxies extracted from the photometric OTELO database \citep{bon19} with a fitted S\'{e}rsic profile \citep{nad20}.
These techniques are (1) the \citet{str01} \ur color algorithm; (2) the LDA machine learning algorithm, which includes both the \ur color and a shape parameter, either the S\'{e}rsic index or the concentration index \citep{kel12}; and (3) a \dnn that uses optical and near-infrared photometry, and shape parameter for objects available in both OTELO and COSMOS catalogs.
We find that a simple, easily trainable \dnn yields a highly accurate classification for ET and LT OTELO galaxies.
Moreover, we apply our \dnn architecture to a set of tabulated COSMOS data with some differences in the photometric bands measured with respect to OTELO, and find that our architecture also performs accurate classification of COSMOS galaxies.
Finally, we use the same \dnn architecture but substituting the S\'{e}rsic index with the concentration index \citep{shi01} for both OTELO and COSMOS datasets.

This paper is organized as follows. 
Section~\ref{sec:method} describes the different techniques used to classify galaxies. 
In Section~\ref{sec:results} we show the results and compare the different techniques. 
Finally, in Section~\ref{sec:conclus} we present our conclusions.


\section{Methodology}\label{sec:method}


The current investigation involves the automatic classification of galaxies into two dichotomous groups, namely ET and LT galaxies, using both photometric measurements and a factor that depends on the shape of the galaxys' light distribution.
Machine learning algorithms for automatic classification parse data and learn how to assign subjects to different classes.
These algorithms require both training and test datasets that consist of labeled data.
The training dataset is used to fit the model parameters, and the test dataset to provide an unbiased assessment of the model performance.
If the algorithm requires tuning the model hyperparameters, such as the number of layers and hidden units in a \dnn architecture, a third labeled dataset called the validation dataset is required to evaluate different model trials (the test dataset must be evaluated only by the final model).
Once the final model architecture is attained, it is trained joining both the training and the validation dataset, and then evaluated using the test dataset.

In this section we present our samples of galaxies extracted from OTELO and COSMOS.
We use the observed photometry and colors, that is, neither $k$ nor extinction corrections were performed.
In order to maximize the sample size while keeping a well-sampled set in redshift, data have been limited in photometric redshift ($z_{phot} \leq \zlim$) but not in flux, thus no cosmological inferences can be performed from our sample.
However, at the end of Sect.~\ref{sec:results} we present a brief analysis of the results obtained for flux-limited samples.
We describe the photometry and the shape factors of these data.
We then present the implementation of the different classification methodologies used: the \ur color, LDA, and \dnn.
Finally, we present the bootstrap procedure that we use to compare the results obtained with these methodologies.

\subsection{OTELO samples}

OTELO is a very deep blind survey performed with the red tunable filter OSIRIS instrument of the 10.4\,m Gran Telescopio Canarias \citep{bon19}.
OTELO data consist of images obtained in 36 adjacent narrow bands (FWHM 12\,\AA) covering a window of 230\,\AA\ around $\lambda = 9175$\,\AA.
The catalog includes ancillary data ranging from X-rays to far infrared.
Point spread function-model photometry and library templates were used for separating stars, AGNs, and galaxies.

The OTELO catalog comprises \numprint{11237} galaxies.
\citet{nad20} matched OTELO with the output from GALAPAGOS2 \citep{hau07, hau13} over high-resolution HST images.
Not all the OTELO galaxies were detected by GALAPAGOS2, which returned a total of 8812 sources.
\citet{nad20} account for automated detection of multiple matches produced by more than one source that lay inside the OTELO's Kron radius in a high-resolution F814W band image.
These latter authors attribute these multiple matches to close companions (gravitationally bounded or in projection), mergers, or resolved parts of the host galaxy. 
In any case, sources with multiple matches were excluded from our analysis because they could affect low-resolution photometry.
Finally, we included further constraints to extract our OTELO samples (see below).

\subsubsection{S\'{e}rsic index and photometry sample}

OTELO uses LePhare templates to fit the SED of galaxies to obtain photometric morphological type classification and redshift estimates.
We used this morphological classification to assign the galaxies to ET and LT classes.
The best model fitting is recorded under the \texttt{MOD\_BEST\_deepN} numerical coded entries in the OTELO catalog \citep{bon19}.
The ET class includes galaxies coded as `1' in the OTELO catalog, which were best fitted by the E/S0 template from \citet{col80}. 
The LT class comprises OTELO galaxies coded from `2' to `10', which were best fitted by different late-type galaxy templates, namely Sbc, Scd, and Irr \citep{col80}, and starburst-class templates from SB1 to SB6 \citep{kin96}.
\citet{bon19} estimate that the fraction of inaccurate SED fittings for the galaxies contained in the OTELO catalog may amount to up to $\sim 4\%$.
Therefore, our results may be affected if there are ET galaxies miscoded differently from `1' in OTELO, or any of the LT galaxies miscoded as `1'.
This could affect, for example, early-type spirals such as Sa galaxies, which are not explicitly included in the OTELO template set.
However, the UV SED for ellipticals and S0 galaxies is completely different from Sa and other LT galaxies.
The OTELO catalog also includes GALEX-UV data that allow us to identify ET galaxies even in the local universe.
Thus, we conclude that recoding the OTELO classification in our galaxy sample as ET and LT classes yields a negligible number of misclassified objects (certainly much less than the OTELO fraction of $\sim4\%$)  and does not affect our results.

The S\'{e}rsic profile is a parametric relationship that expresses the intensity of a galaxy as a function of the distance from its center:

\begin{equation}\label{eq:sersic}
  I(R) = I_e e^{-b \big[ \big( \frac{R}{R_e} \big)^{1/n} -1 \big]},
\end{equation}

\noindent
where $I$ is the intensity at a distance $R$ from the galaxy center, $R_e$ is the half-light radius, $I_e$ is the intensity at radius $R_e$, $b \ (\sim 2n-1/3)$ is a scale factor, and $n$ is the S\'{e}rsic index.
S\'{e}rsic profiles have been employed for galaxy classification \citep[\eg][]{kel12, vik15}.
This index provides a geometrical description of the galaxy concentration; for a S\'{e}rsic index $n = 4$ we obtain the de Vaucouleurs profile typical of elliptical galaxies, while setting $n = 1$ gives the exponential profile describing spiral galaxies.


Our OTELO S\'{e}rsic index and photometry (OTELO \ps) sample consists of \ngal galaxies at redshifts $z \leq \zlim$ extracted from the OTELO catalog (listed under the \texttt{Z\_BEST\_deepN} code).
The sample includes $ugriz$ optical photometry from the Canada-France-Hawaii Telescope Legacy Survey\footnote{\url{http://www.cfht.hawaii.edu/Science/CFHTLS/}} (CFHTLS), $JHKs$ near-infrared photometry from the WIRcam Deep Survey\footnote{\url{https://www.cadc-ccda.hia-iha.nrc-cnrc.gc.ca/en/cfht/wirds.html}} (WIRDS), and S\'{e}rsic index estimates obtained using GALAPAGOS2/GALFIT \citep{hau13, pen02, pen10} on the HST-ACS publicly available data in the F814W band. 
The sample comprises only galaxies with S\'{e}rsic indexes between $n = 0.22$ and $n = 7.9$; S\'{e}rsic indexes out of this range are not reliable because of an artificial limit imposed by the S\'{e}rsic-profile-fitting algorithm in GALAPAGOS2 \citep{hau07}. 
Besides, the sample does not include galaxies with S\'{e}rsic index values less than three times their estimate errors.
For a detailed description of the S\'{e}rsic-profile-fitting process we refer to \citet{nad20}.

\begin{figure*}
        \centering
        \includegraphics[width=0.5\textwidth]{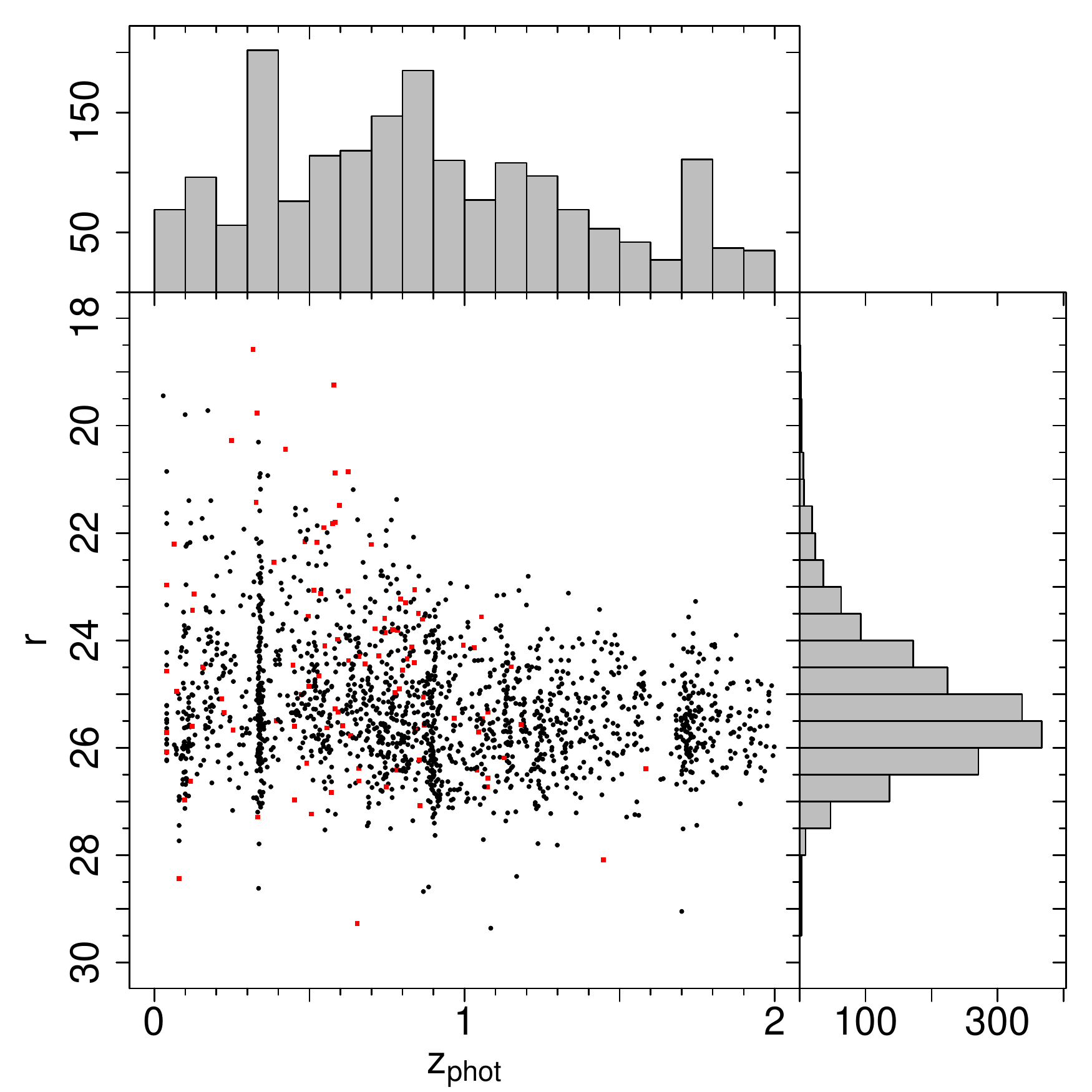}
        \caption{Comparative distribution of brightnesses in the $r$ band and photometric redshifts for the \ngal galaxies in the OTELO \ps sample. \emph{Bottom left panel:}  $r$ magnitude vs.\ photometric redshift $z_{phot}$ plot shows the galaxies in the sample, differentiating between LT galaxies (black circles) and ET galaxies (red squares). \emph{Top panel:}  \ps sample photometric redshift distribution. \emph{Right panel:}  \ps sample $r$ magnitude distribution.} \label{fig:ps}
\end{figure*}

Figure \ref{fig:ps} shows the sample distributions of magnitudes in the $r$ band and photometric redshifts extracted from the OTELO catalog.
We note that the sample is not limited in flux, and therefore it is not a complete sample in the volume defined by the redshift limit $z_{phot} \leq \zlim$ (see the discussion about magnitude-limited samples below).
The redshift distribution presents concentrations at redshifts 0.04, 0.11, 0.34, 0.90, and 1.72 superimposed onto a bell-like distribution with a maximum around $z_{phot} \approx 0.8$ and a strong decay from $z_{phot} \approx 1.3$.
The photometric data are incomplete, which affects the available number of galaxies for those classification procedures that cannot effectively manage missing data.

The sample is randomly divided in a training set (70\% of the available galaxies) used for the algorithm training, and a test set (30\%) used to yield an unbiased estimate of the efficiency of the model.
Choosing the proportions of training and test sample sizes depends on a balance between the model performance and the variance in the estimates of the statistical parameters (in our case the accuracy, \sensitivity and \specificity, as explained below).
Rule-of-thumb proportions often used in machine learning are 90:10 (\ie 90\% training, 10\% testing), 80:20 (inspired by the Pareto principle), and 70:30 (our choice).
In our case, the 70:30 proportion is justified because it fulfills the \emph{large enough sample condition} (another rule of thumb) that the sample size must be at least 30 to ensure that the conditions of the central limit theorem are met. 
Thus, the number of expected ET galaxies in the \ps test sample is: $N_{gal} p_{et} p_{test} \approx \FPeval\result{round(\soln * 0.054 * 0.3, 0)}\numprint{\result}$, where $N_{gal} = \soln$ is the sample size, $p_{et} = \FPeval\result{round(\net / \soln, 3)}\numprint{\result}$ is the proportion of ET galaxies in the \ps sample, and $p_{test} = 0.3$ is the proportion of galaxies in the test sample.

\subsubsection{Concentration and photometry sample}

 
 The concentration is widely used to differentiate ET from LT galaxies.
 Concentration provides a direct measurement of the intensity distribution in the image of a galaxy.
For that reason, the concentration is easier to obtain than the S\'{e}rsic index, which requires fitting several parameters to the S\'{e}rsic profile.

\begin{figure*}
        \centering
        \includegraphics[width=0.5\textwidth]{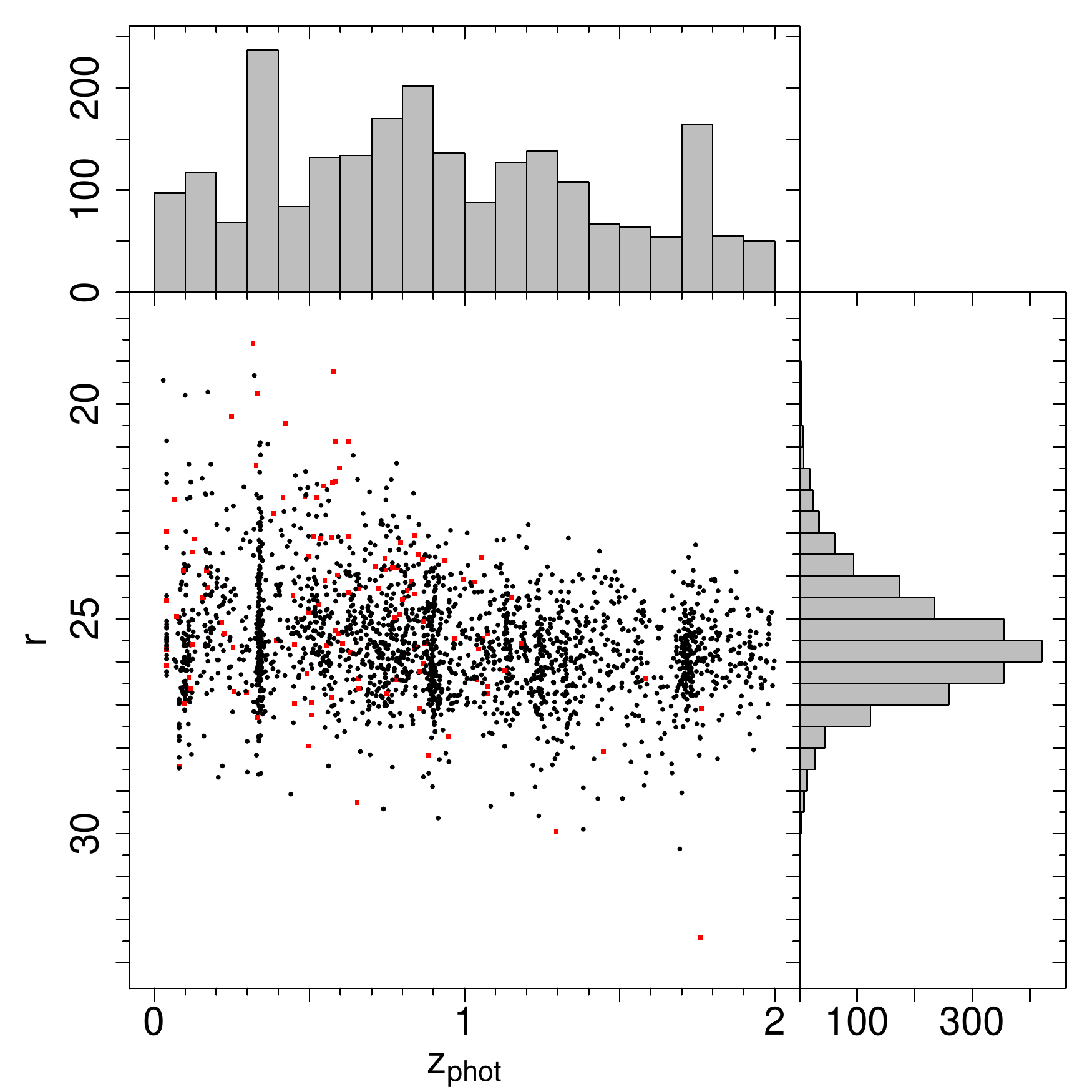}
        \caption{As Fig. \ref{fig:ps} but for the \ngalc galaxies in the OTELO \pc sample.} \label{fig:pc}
        \end{figure*}

Here we use the definition \citep{ber00,sca07}:
\begin{equation}
        C = 5 \log_{10} \! \left( \frac{r_{80}}{r_{20}} \right),
\end{equation}
where $r_{80}$ and $r_{20}$ are the 80\% and 20\% light Petrosian radii, respectively, obtained from the HST F814W band images.
We chose the F814W band concentration for compatibility with COSMOS \citep{sca07}. 
The data were limited to a redshift $z \leq \zlim$. 
The final OTELO concentration and photometry (OTELO \pc) sample consists of \ngalc galaxies, with \netc\ classified as ET and \nltc\ as LT.
Figure \ref{fig:pc} shows the sample distributions of magnitudes in the $r$ band and photometric redshifts, which is similar to the case of the \ps sample discussed above.
The \pc sample was also divided in two subsamples: a training subsample containing \ntrainc (70\%) of the objects, and a test subsample with \ntestc (30\%) of the galaxies.

\subsection{COSMOS samples}
We expect that our \dnn architecture can be applied to galaxy classification in other databases.  
Therefore, we checked its reliability using two COSMOS enhanced data products: the \emph{Zurich Structure \& Morphology Catalog v1.0} \citep[\zurich,][]{sca07, sar07} and the \emph{COSMOS photometric redshifts v1.5} \citep[\photoz,][]{ilb09}. 
Those catalogs have \numprint{131532} and \numprint{385065} entries, respectively.
We merged both databases, obtaining \numprint{128442} matches, from which we chose a sample of galaxies with S\'{e}rsic indexes estimates in the range $0.2 < n < 8.8$,  and another sample with the same concentration radii used in OTELO.
Both samples are limited to redshifts $z<2$ and include photometry in the CFHT~$u$, Subaru~$BVgriz$, UKIRT~$J$ and CFHT~$K$ bands, along with classification entries.
Thus, the galaxy records included all the available data from the \photoz bands except the CFGT $i^\prime$ magnitudes (we chose the Subaru $i$ band also included in the catalog).

The resulting COSMOS S\'{e}rsic index and photometry (COSMOS \ps) sample consists of \ngalC galaxies, \numprint{\nltC} of which had been classified as LT and \netC\ as ET.
With such a large number of galaxies, we can limit the training set to \ntrainC\ galaxies (a fraction of approximately \FPeval\result{round(100*\soltrainC/\solnC,0)}\numprint{\result}\% of the sample), and rise the fraction of the testing set up to \ntestC galaxies (approximately \FPeval\result{round(100*\soltestC/\solnC,0)}\numprint{\result}\% of the sample) in order to reduce the variance of the results.
Analogously, the COSMOS concentration and photometry (COSMOS \pc) sample consists of \ngalCc galaxies, distributed in \numprint{\nltCc}\ LT and \netCc\ ET. 
We set the corresponding training and testing sets to \ntrainCc and \ntestCc galaxies, respectively.

\begin{table}[!t]
\caption[]{One color separation.}\label{tab:colors}
{\centering
\begin{tabular}{lr@{$\,\pm\,$}lrr@{$\,\pm\,$}l}
  \hline
  \noalign{\smallskip}
Color & \multicolumn{2}{c}{Accuracy} & N$_{test}$ & \multicolumn{2}{c}{Separation} \\ 
  \noalign{\smallskip}
  \hline
  \noalign{\smallskip} 
  $u-J$  & 0.96 & 0.01 & 518 & \phantom{0}4.1 & 0.2 \\
  $u-i$  & 0.96 & 0.01 & 536 & 2.8 & 0.3  \\ 
  $u-r$  & 0.96 & 0.02 & 536 & 2.0 & 0.2  \\
  $u-H$  & 0.95 & 0.01 & 510 & \multicolumn{2}{c}{Baseline} \\
  $g-J$  & 0.94 & 0.02 & 527 & \multicolumn{2}{c}{Baseline} \\
  $g-i$  & 0.94 & 0.01 & 548 & \multicolumn{2}{c}{Baseline} \\
  \noalign{\smallskip}
  \hline
  \noalign{\smallskip}
\end{tabular}
\par } 
\footnotesize{\textbf{Notes:} 
  Column 1: tested color. 
  Column 2: mean accuracy (proportion of galaxies correctly classified) on 100 random test sets; we note that the colors are sorted from the highest to the lowest accuracy score.
  Column 3: sample size included in each test set; differences in the sample size between different colors are due to missing data.
  Column 4: color discriminant value or \emph{Baseline} if the accuracy score is not statistically different from the baseline classification.} 
\end{table}

\subsection{Classification procedures}
We used a classification baseline and three classification methods for the OTELO samples.
The baseline consists in classifying all the galaxies into the most frequent morphological group.
Any classification by a more sophisticated method should improve the baseline accuracy.
For the COSMOS samples we only used the classification baseline and the \dnn architecture developed for the OTELO samples, as we were interested only in probing this architecture. 

\subsubsection{Color classification}

The first classification method uses a color discriminant.
After testing several colors, we focus on the \ur color as proposed by \citet{str01}. 
These authors use a simple color discriminant such that any galaxy with \ur color redder than \urplane is classified as ET, and LT if \(\ur < \urplane.\)
This method was applied only to both \ps and \pc samples drawn from OTELO.
We also investigated other possible color discriminants that will be presented later.
Data records with missing \ur colors were disregarded, reducing the \ps sample to \nurgal galaxies and the \pc sample to \nurgalc.

\subsubsection{Linear discriminant analysis}

The second classification method is LDA.
The aim of LDA is to find a linear combination of features which separates different classes of objects.
These features are interpreted as a hyperplane normal to the input feature vectors.
We note that the \citet{str01} \ur color separation method can be regarded as a LDA which defines the \(\ur = \urplane\) plane normal to $u-g$ and $g-r$ vectors. 
As in the previous method, LDA was only applied to \ps and \pc OTELO samples, and data records with missing \ur colors were disregarded.

\begin{table}[!t]
\caption{Color \ur confusion matrix.}\label{tab:urconf} 
\centering
\begin{tabular}{crrrr}
  \hline
  \noalign{\smallskip}
  &       & \multicolumn{2}{c}{OTELO} \\
  \cline{3-4}
  \noalign{\smallskip}
  &    & ET & LT & \textbf{Total} \\ 
  \noalign{\smallskip}
  \hline
  \noalign{\smallskip}
  \multirow{2}{*}{Color} & ET  & 25 &  20 &  \textbf{45} \\ 
                         & LT  &  2 & 489 &  \textbf{491} \\
  & \textbf{Total} & \textbf{27} & \textbf{509} & \textbf{536} \\ 
  \noalign{\smallskip}
  \hline
\end{tabular}
\end{table}



Two problems with machine learning techniques are the management of missing data and the curse of dimensionality.
Missing data (\eg\ a photometric band) usually results in removing objects with incomplete records from the dataset.
The curse of dimensionality appears because increasing the number of variables in a classification scheme means that the volume of the space increases very quickly and therefore the data become sparse and difficult to group.
The curse of dimensionality can be mitigated by dimensionality reduction techniques such as principal component analysis (PCA), but dimensionality reduction may introduce unwanted effects \citep[data loss, nonlinear relations between variables, and the number of components to be kept]{car01,shl14}  that tend to blur differences between the groups.
Alternative methodologies to deal with these problems are under development, for example by \citet{cai18} who introduce an adaptive classifier to cope with both missing data and the curse of dimensionality for high-dimensional LDA.
To avoid these problems, we chose to limit our LDA model to the S\'{e}rsic index and the single highly discriminant \ur color, as it has been already addressed in the galaxy classification literature \citep[\eg][]{kel12, vik15}.

\begin{table*}[t]
\caption{Comparison of classification methods for \ps samples.}\label{tab:compar}
\centering
\begin{tabular}{llllllllllr}
  \hline
  \noalign{\smallskip}
  \multirow{2}{*}{Database} & \multirow{2}{*}{Method} & \multicolumn{2}{c}{Accuracy$\,^a$} & & \multicolumn{2}{c}{\sensitivity$\!^a$} & & \multicolumn{2}{c}{\specificity$\!^a$} & \multicolumn{1}{c}{Sample} \\
  \cline{3-4}
  \cline{6-7}
  \cline{9-10}
  \noalign{\smallskip}
 & & Mean & Error & & Mean & Error & & Mean & Error & \multicolumn{1}{c}{size} \\ 
  \noalign{\smallskip}
  \hline
  \noalign{\smallskip}
  OTELO & Baseline  & \FPupn\result{\plt{} 100 swap / 3 round}\result & 0.006 && 0 & ~\dots && 1 & ~\dots &  \ngal \\ 
  OTELO & \ur color & 0.96 & 0.02 && 0.8 & 0.3 && 0.97 & 0.02 & 536 \\ 
  OTELO & LDA       & 0.970 & 0.008 && 0.80 & 0.08 && 0.979 &  0.007 & 536 \\
  OTELO & \dnn       & 0.985 & 0.007 && 0.84 &  0.09 && 0.993 & 0.006 & 551 \\
    \noalign{\smallskip}
  COSMOS & Baseline  & \FPupn\result{\pltC{} 100 swap / 3 round}\result & 0.002 && 0 & ~\dots && 1 & ~\dots &  \ngalC \\ 
  COSMOS & \dnn       & 0.967 & 0.002 && 0.91 & 0.03 && 0.979 & 0.005 & \FPeval\result{round(\solnC-\soltrainC, 0)}\numprint{\result} \\
 
  \noalign{\smallskip}
  \hline
  \noalign{\smallskip}
  \multicolumn{4}{l}{\footnotesize{$^a$ On 100 bootstrap runs.}}
\end{tabular}
\end{table*}

\subsubsection{Deep neural network}
The third method of classification involves a \dnn.
The sample was analyzed using the Keras library for deep learning.
Keras is a high-level neural network application programming interface (API) written in Python under GitHub license.
Currently, Keras is available for both Python and R computer languages \citep{cho17a, cho17b}.
In astronomy, Keras has already been used for image classification of galaxy morphologies \citep{per18, dom18} and spectral classification and redshift estimates of quasars \citep{bus18}, and is included in the astroNN package.\footnote{\url{https://astronn.readthedocs.io/en/latest/}}

As in the other methods, we use a training and a test set to teach and check the \dnn model,
respectively.
The difference from the other methods is that the structure of their learning discriminant function is predetermined, while the \dnn architecture should be tuned on the fly.
To achieve this goal, we split the training set in the OTELO samples in (i) a \teach (80\% of the original training set), and (ii) a validation set (the remaining 20\%).
Compared with OTELO, the COSMOS samples consist of many more galaxies. 
Therefore, we limited the number of the training sets to \ntrainC for the COSMOS \ps sample, and \ntrainCc for the \pc sample, and conserved the respective \teach and validation set proportions.
We use the \teach to tune the \dnn model, and the validation set to check the loss and accuracy functions that describe the \dnn classification capability.
Once we have achieved a satisfactory result, the \dnn architecture has been optimized to classify the validation set, but the performance may be different for other datasets.
To generalize the result, we use the whole original training set to retrain the tuned \dnn model, and we then classify the test set galaxies.
Therefore, the test set galaxies were used neither to train nor to fine tune the \dnn model, but only to evaluate the \dnn performance.

An appealing feature of \dnns is the easiness to deal with missing data.
In practice, it is enough to substitute the missing values in each normalized variable by zeros to cancel their products on the network weights.
The \dnn then deals with missing values as if they do not carry any useful information and will ignore them.
Of course, it is better if there are not missing values, but \dnns allow the user to treat them without the need of dropping data entries or estimating missing values from other variables.

\citet{bar19} provides a succinct description of \dnns, and a complete explanation of Keras elements can be found in \citet{cho17a} and \citet{cho17b}.
Tuning a \dnn is a trial-and-error procedure aimed to find an appropriate architecture and setup.
As the numbers of input variables, units, and layers increase, the \dnn tends to overfit if the training set is small.
For this reason, we kept our \dnn model as simple as possible whilst obtaining a high-accuracy classification.

We use standard layers and functions  for our
model that are already available from Keras.
For the interested reader, our \dnn architecture consists of two dense layers of 64 units each with \emph{rectified linear unit} (ReLU) activations, and an output dense layer of a single unit with \emph{sigmoid} activation.
The model was compiled using an iterative gradient descendent RMSprop optimizer, a binary-cross-entropy loss function, and accuracy metrics.
%
We kept the default values for the Keras RMSprop optimizer, \ie a learning rate of 0.001 and a weight parameter for previous batches of $\rho = 0.9$.
These values are appropriate for most \dnn problems, and moderate changes do not affect the results. 
We set the number of training epochs to avoid overfitting, and the training batch sizes to appropriate values for the number of records in the DNN training sample in each case.

\subsubsection{Bootstrap}
We used bootstrap \citep[\eg][]{efr93, chi18} to obtain reliable statistics that describe the performance of each classification technique.
Bootstrapping is a widely used nonparametric methodology for evaluating the distribution of a statistic using random resampling with replacement. 
Thus, we calculated the classification accuracy and other classification statistics through 100 runs for the \ur color, LDA, and DNN methods.
For each run, we also divided the bootstrap random sample in a training set (70\%) and a test set (30\%).

\section{Results}\label{sec:results}

To determine a minimal set of attributes that are able to classify between ET and LT galaxies, we focus on two directly observable characteristics: photometry and shape.
Results obtained in previous studies were limited to nearby galaxies.
Thus, \citet{str01} used photometry from \numprint{147920} SDSS galaxies with magnitude $g^* \leq 21$ and redshifts $z \lesssim 0.4$ to build a binary classification model based in the $\ur = \urplane$ discriminant color, which they tested on a sample of 287 galaxies visually labeled as ET or LT, recovering 94 out of 117 (\FPeval\result{round(94 / 117 * 100, 0)}\result\%) ET, and 112 out of 170 (\FPeval\result{round(112 / 170 * 100, 0)}\result\%) LT galaxies.
\citet{den13} used a sample of \numprint{233669} SDSS-III DR8 galaxies with redshifts $0.01 < z < 0.25$ and report a concentration index discriminant to separate ET from LT galaxies in the $r$-band that achieved an accuracy of \FPupn\result{233669 46768 178557 + /} \FPupn\erresult{233669 \result{} \result{} 1 - / / 2 swap root}\FPeval\result{round(\result{} * 100, 2)}\result\ $\pm$ \FPeval\erresult{round(\erresult{} * 100, 2)}\erresult. 
\citet{vik15} used both the \ur color and the S\'{e}rsic index in the $r$-band to classify a sample of 142 nearby ($z < 0.01$) galaxies, dividing the \ur versus $n_r$ plane in quadrants; most ET galaxies were located at the $\ur > 2.3$ and $n_r >2.5$ quadrant (28 out of 34, \ie \FPeval\result{round(28 / 34 * 100, 0)}\result\% ETs were correctly classified).

\begin{table}[t]
\caption{LDA confusion matrix}\label{tab:ldaconf}
\centering
\begin{tabular}{llrrr}
  \hline
  \noalign{\smallskip}
 &        & \multicolumn{2}{c}{OTELO} \\
  \cline{3-4}
  \noalign{\smallskip}
 & & ET & LT & \textbf{Total} \\ 
  \noalign{\smallskip}
  \hline
  \noalign{\smallskip}
  \multirow{2}{*}{LDA} & ET & 20 & 8  & \textbf{28} \\ 
                       & LT &  7 & 501 & \textbf{508} \\ 
  & \textbf{Total} & \textbf{27} & \textbf{509} & \textbf{536}\\
  \noalign{\smallskip}
  \hline
\end{tabular}
\end{table}


\subsection{\sp samples}

\subsubsection{Baseline classification}
The baseline classification is the simplest classification method.
It assigns all the samples to the most frequent class.
This classification is helpful for determining a baseline performance that is used as a benchmark for other classification methods.
For this task, we selected all the galaxies in our OTELO SP sample.
In total, there are \ngal galaxy records, \net\ of them classified as ET galaxies (\FPeval\result{round(\pet, 1)}$\approx \result$\%), and \nlt\ as LT galaxies (\FPeval{\result}{round(\plt, 1)}$\approx \result$\%). 
The two groups are unevenly balanced, which results in the baseline classification achieving a high overall accuracy of \result\%, which should be exceeded by any other classification method.

\subsubsection{Color classification}
A preliminary study was performed to decipher which colors yield a split between ET and LT galaxies that outperforms the baseline.
Table \ref{tab:colors} shows several examples of the measured accuracy for selecting appropriate single color discriminants.
We note that several colors did not perform better than the baseline classification (\FPupn\result{\plt{} 1 round}\result\%), but those involving the $u$ and a red band usually yield the most accurate results.
Both $u-J$ and $u-i$ colors perform marginally better than \ur, although $u-J$ has a larger number of missing records.
We present the rest of the color analysis based on the \ur color in order for ease of direct comparison with the report of \citet{str01}. 

Table~\ref{tab:urconf} shows an example of the confusion matrix for a single \ur color  bootstrap run, yielding an accuracy of $0.959 \pm 0.009.$
Table~\ref{tab:compar} shows the Accuracy, \sensitivity, and \specificity for the different databases and classification methods used in this paper, obtained through the bootstrap procedure.
The \sensitivity and \specificity both indicate the proportion of ET and LT galaxies, respectively, recovered through the classification procedure.
For the \ur color, the statistics yield an average Accuracy of $0.96 \pm 0.02$, a \sensitivity of $0.8 \pm 0.3$, and a \specificity of $0.97 \pm 0.02$.
The \sensitivity is the least precise of all the statistics in all the samples because of the relatively low number of ET galaxies. 

\begin{table}[t]
\caption{OTELO \dnn confusion matrix}\label{tab:dnn}
\centering
\begin{tabular}{llrrr}
  \hline
  \noalign{\smallskip}
 &        & \multicolumn{2}{c}{OTELO} \\
  \cline{3-4}
  \noalign{\smallskip}
 & & ET & LT & \textbf{Total} \\ 
  \noalign{\smallskip}
  \hline
  \noalign{\smallskip}
  \multirow{2}{*}{\dnn} & ET & 27 &   4 &  31 \\ 
                                            &LT  &  4 & 516 & 520 \\ 
                 & \textbf{Total}    & 31 & 520 & 551 \\
   \noalign{\smallskip}
   \hline
\end{tabular}
\end{table}



\begin{figure*}
   \centering
   \includegraphics[width=0.5\textwidth]{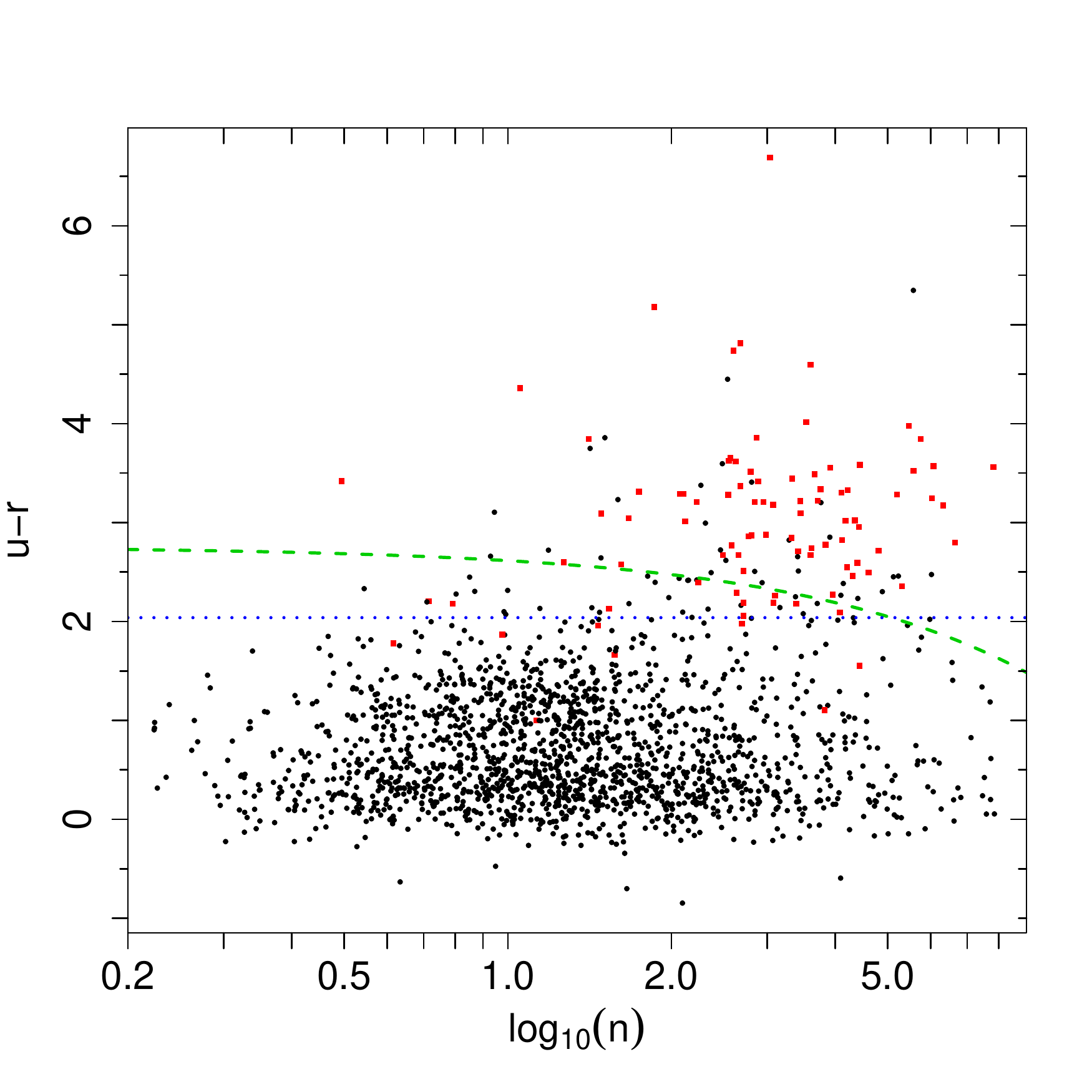}
   \caption{Classification for the OTELO \ps sample of \ngal galaxies through the \ur color and LDA algorithms. The \ur color vs.\ logarithm of the S\'{e}rsic index $n$ plot shows the original morphological type classification in the OTELO catalog reduced to LT galaxies (black circles) and ET galaxies (red squares). The dotted blue line indicates the \ur color separation, and the dashed green line the LDA separation by the \ur color and the S\'{e}rsic index. The logarithmic scale for the S\'{e}rsic index makes the comparison with the concentration in Fig.~\ref{fig:lda_conc} (which is already a logarithmic quantity) easier, but it bends the LDA line. The reader should take into account that this is not a flux limited sample.}\label{fig:lda}%
   \end{figure*}

\begin{figure*}
   \centering
   \includegraphics[width=0.5\textwidth]{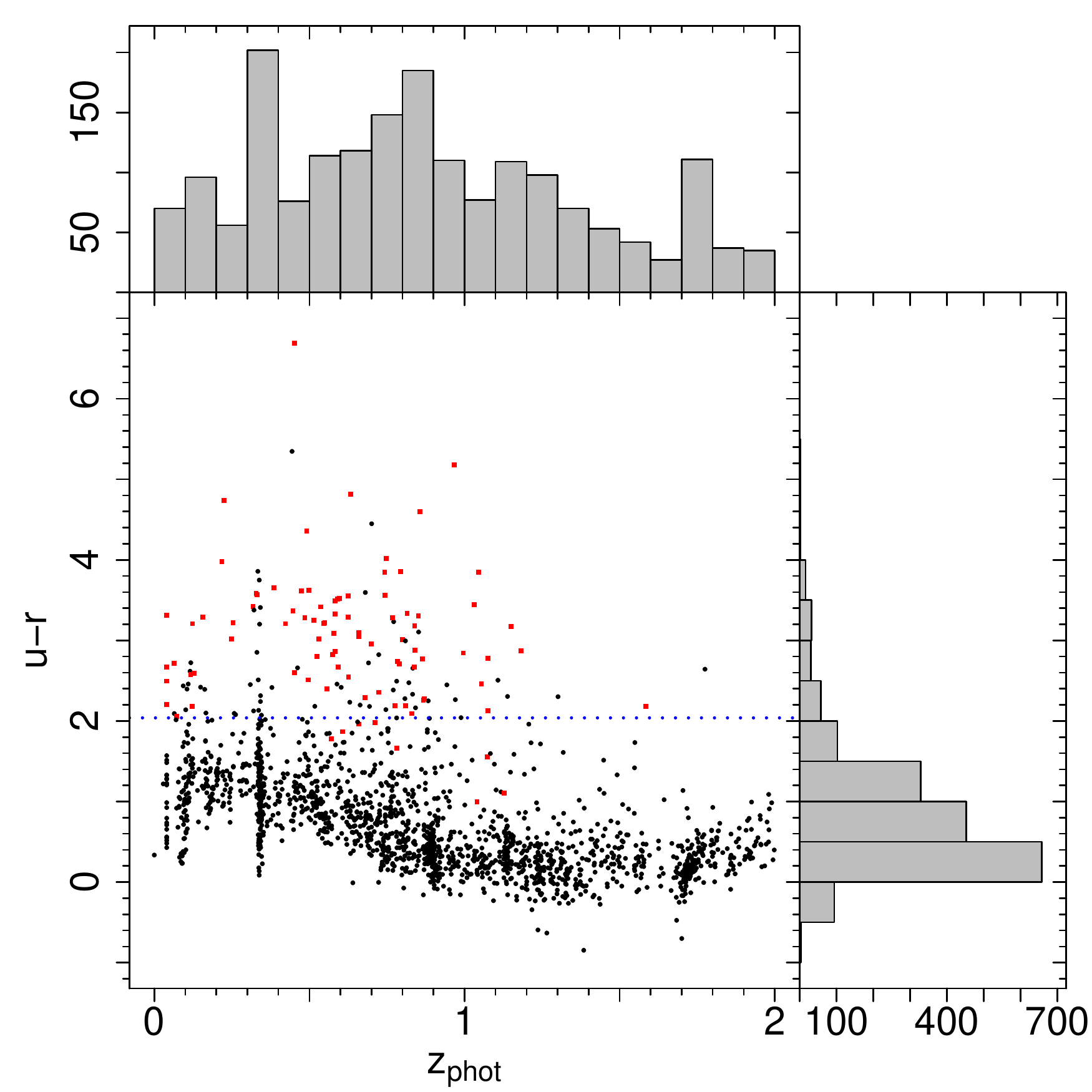}
   \caption{Comparative distribution of the \ur color and redshifts for the \ngal galaxies in the OTELO \ps sample. \emph{Bottom left panel:}  \ur color vs.\ photometric redshift $z_{phot}$ plot shows the original morphological type classification in the OTELO catalog reduced to LT galaxies (black circles) and ET galaxies (red squares). The dotted blue line indicates the \ur color separation.  \emph{Top panel:}  \ps sample photometric redshift distribution. \emph{Right panel:}  \ps sample \ur color distribution.} \label{fig:ur_z_sp}%
   \end{figure*}

Bootstrap yields a \ur color discriminant for ET and LT separation of $2.0 \pm 0.2$, as shown in Fig.~\ref{fig:lda}.
The agreement with \citet{str01}, $\ur = \urplane$ (no error estimate is provided by these authors), is remarkable considering that the galaxies studied by these authors have redshifts in the interval $0 < z \leq 0.4$ while our sample expands to $z \leq 2$. 
Figure \ref{fig:ur_z_sp} shows the \ur color distribution as a function of the redshift for the OTELO SP sample.
It is worth noting that LT galaxies in this sample tend to be bluer at redshifts $z \lesssim 0.5$, possibly due to an enhanced star-forming activity as also pointed out by \citeauthor{str01} 
This feature, along with the scarcity of ET galaxies at $z > 1$ (about 13\% of all the ET galaxies in the OTELO SP sample), justifies the agreement between \citeauthor{str01} results and ours despite the redshift differences.

\begin{figure*}
   \centering
   \includegraphics[width=0.5\textwidth, trim = 0 70 0 0, clip]{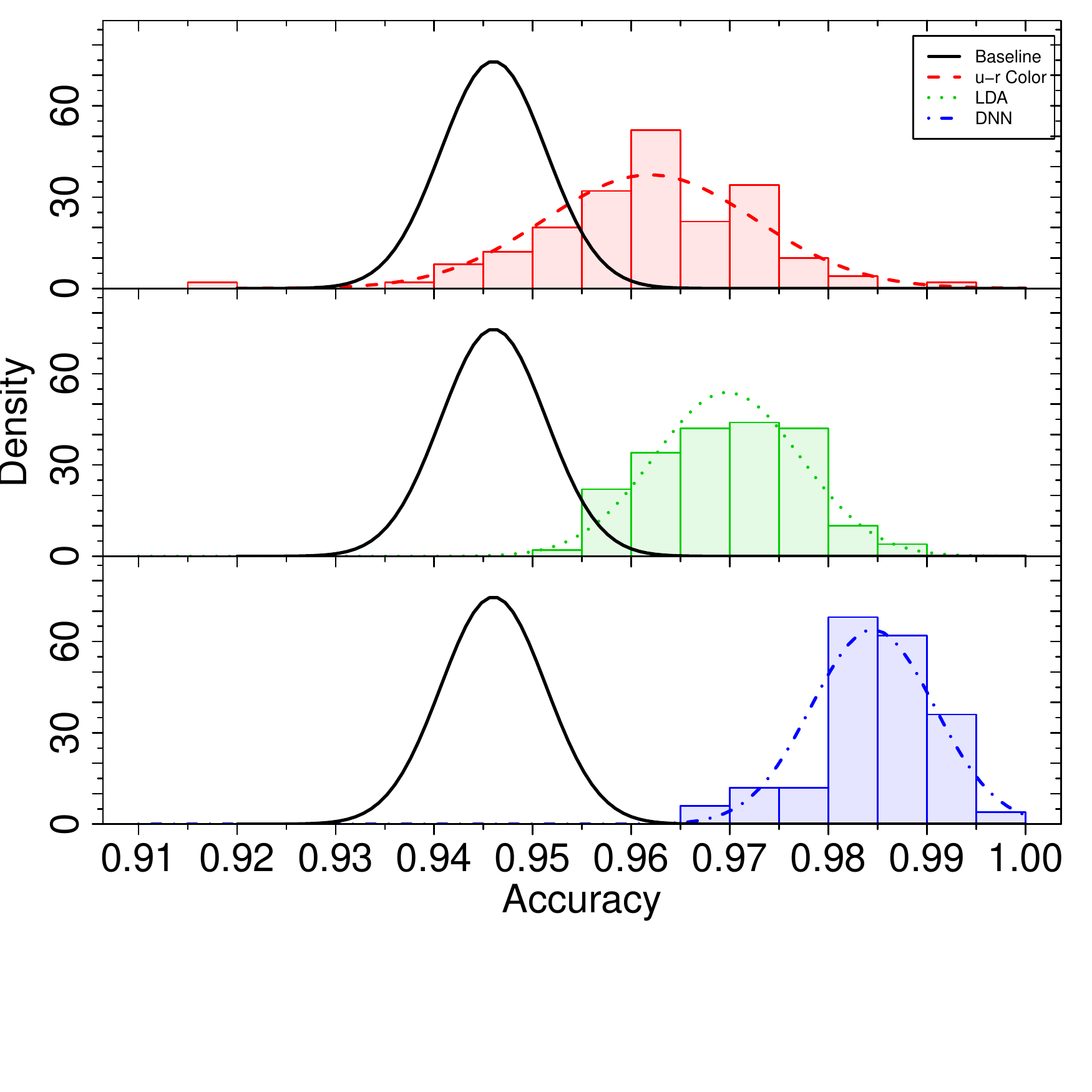}
   \caption{Histogram of accuracies for 100 galaxy classification bootstrap runs for the OTELO \ps sample.
   The solid black lines correspond to the baseline accuracy distribution of the whole sample of \ngal OTELO \ps galaxies.
   The red histogram and the Gaussian fit represented by a dashed line show the \ur color accuracy distribution obtained from 100 bootstrap runs.
   As for the \ur color, the green histogram and the dotted line show the LDA accuracy distribution.
   Similarly, the blue histogram and the dash-dotted line shows the \dnn accuracy distribution.}\label{fig:dnn_acc}%
    \end{figure*}

The distribution of the bootstrap Accuracy for the \ur color classification is shown in the upper panel of Fig.~\ref{fig:dnn_acc}.
Most of the \ur color accuracies are larger than the baseline, but the two extreme bootstrap runs with accuracies lying in the 0.915--0.92 interval fail to detect any ET galaxy.

\begin{figure*}
        \centering
        \includegraphics[width=0.5\textwidth, clip]{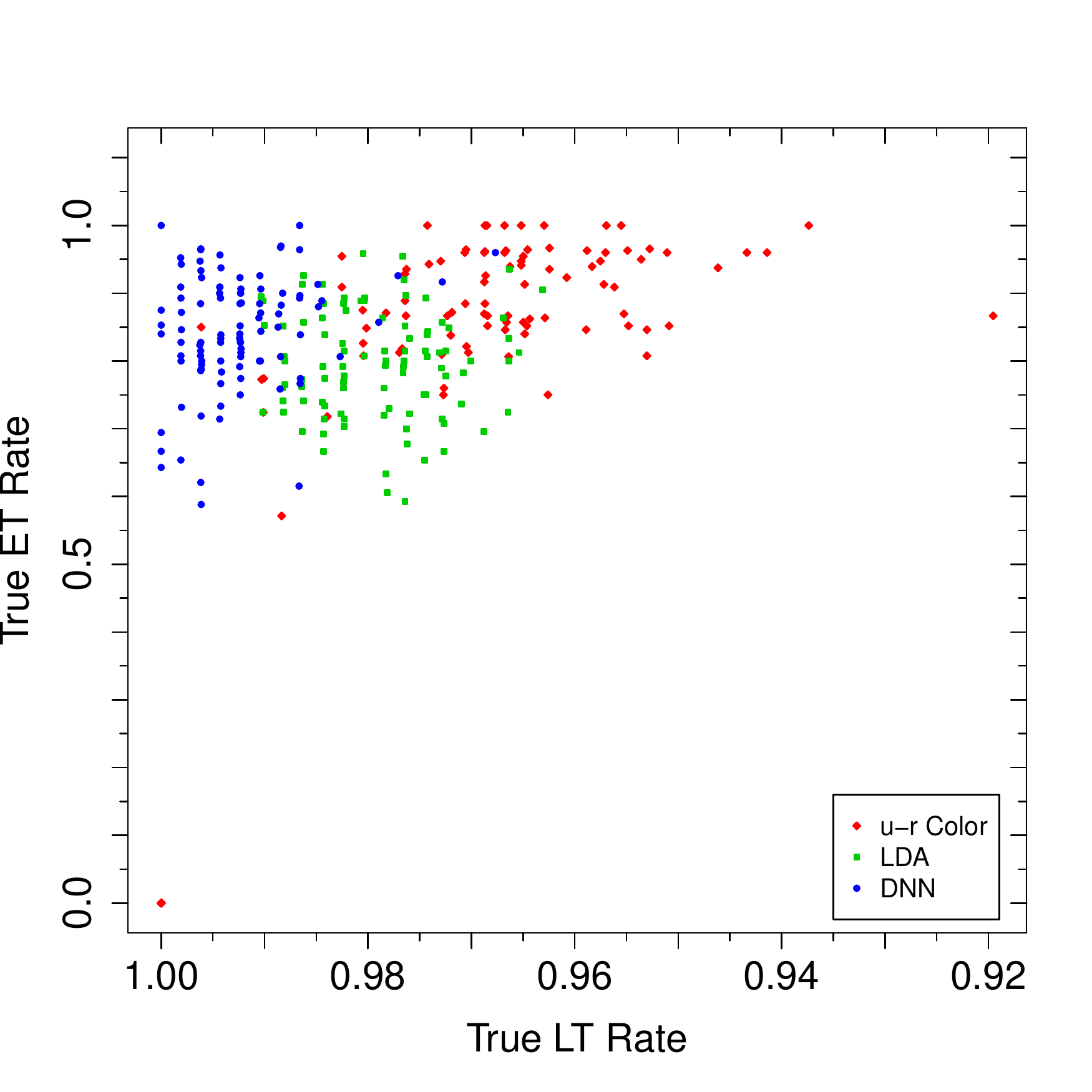}
        \caption{\sensitivity vs. \specificity values for 100 bootstrap runs in each classification method on the OTELO \ps sample.
        The closer to the upper-left corner, the best classification result.
        \dnn runs yield consistently the best classifications.}\label{fig:spvsse}%
\end{figure*}

The \sensitivity and \specificity are analogous to the \emph{True Positive Rate} and \emph{False Positive Rate} ($= 1 - $ \specificity) statistics.
These statistics are used in receiver operating characteristic (ROC) curves to represent the ability to discriminate between two groups as a function of a variable threshold, usually the likelihood of the classification \citep[\eg][]{bar19}.
Figure~\ref{fig:spvsse} shows the distribution of bootstrap values in the \sensitivity versus \specificity plane.
Every point in this figure corresponds to the 50\% probability threshold of the ROC curve (not shown) for each bootstrap run.
The closer the point to the top-right corner, the better the classification.
The data point located at \specificity = 1, \sensitivity = 0 corresponds to the two \ur color bootstrap runs that failed to detect any ET galaxy.
Below, for the LDA and \dnn classification methods, we increase the number of predictor variables used to enhance the distinction between ET and LT galaxies.

\begin{table*}[t]
\caption{OTELO \dnn missmatches}\label{tab:missmatch}
{\centering
\begin{tabular}{rccccccccr}
  \hline
  \noalign{\smallskip}
 \multicolumn{1}{c}{\multirow{2}{*}{ID}	} & \multirow{2}{*}{OTELO} & \multirow{2}{*}{\dnn}	& \multirow{2}{*}{Prob.}  & \multirow{2}{*}{$z_{phot}$} & \multirow{2}{*}{n} & \multirow{2}{*}{C} & \multirow{2}{*}{Elong.} & \multicolumn{1}{c}{Size} \\
    &       &           &               &        &            &   &
       & \multicolumn{1}{c}{(px$^2$)} \\ 
  \hline
  \noalign{\smallskip}

   267	&	LT	&	ET r	&	0.32 	& 0.77 & 1.59 & 3.55 & 1.88 &  259 \\ 
   496	&   LT  &	ET v	&	0.12	    & 0.70 & 2.54 & 3.54 & 1.22 &  231 \\
  1895	&	ET	&	LT r	&	1.00	    & 0.04 & 0.72 & 3.47 & 1.22 &   61 \\ 
  2818	&   LT  &	ET v	&	0.12    & 0.72 & 3.29 & 3.32 & 1.39 &  255 \\  
  3680	&   ET  &	LT v	&	1.00    & 1.45 & 0.36 & 2.14 & 1.49 &   27 \\
  4010	&   LT  &	ET v	&	0.08	    & 0.33 & 3.91 & 4.29 & 1.09 & 1002 \\ 
  4923	&   ET	&	LT v	&	1.00    & 0.08 & 0.41 & 1.12 & 1.04 & 55 \\ 
 10207	&   ET	&	LT v	&	1.00	    & 0.12 & 1.62 & 2.31 & 1.99 & 21 \\
  
  \noalign{\smallskip}
  \hline
  \noalign{\smallskip}
\end{tabular}
\par }
	\footnotesize{\textbf{Notes:} 
	Column 1: galaxy identifier in the OTELO catalog.
	Column 2: OTELO classification.
	Column 3: \dnn classification and a letter code to indicate reject (r) or validate (v); this classification after visual inspection by three of the authors.
	Column 4: \dnn classification likelihood, closer to zero for ET instances and closer to one for LT.
	Columns 5, 6, and 7: the photometric redshift, the S\'{e}rsic index, and the concentration value, respectively.
	Column 8: shows the elongation, that is, the ratio between the major and minor axes of the galaxy image as calculated by SExtractor \citep{ber96}.
	Column 9: shows the area in pixels of the HST Advanced Camera for surveys (scale 0.03 arcsec/px).}
\end{table*}

\subsubsection{Linear discriminant analysis classification}
Although the use of colors is an improvement on the baseline classification, and the \ur plane method is very easy to implement, we dispose of additional data in order to aim for more powerful classification techniques.
In particular, it will be very helpful to include a parameter associated with the galaxy morphology that can be inferred from optical or near-infrared observations. 
The S\'{e}rsic profile in eq.~(\ref{eq:sersic}) describes the intensity of a galaxy as a function of the distance from its center regardless of the galaxy colors, and thus can be useful for our purpose.



The dataset combining \ur colors and S\'{e}rsic indexes has been probed using linear discriminant analysis.
The sample with complete records consisted of \nurgal galaxies which have been split in a training group of \FPupn{\result}{0.7 \nur{} * 0 round}\result\ and a test group of \FPupn{\nurtest}{\result{} \nur{} - 0 round}\nurtest.
Figure~\ref{fig:lda} shows the LDA separation in the \ur color versus\ S\'{e}rsic Index $n$ plane for the test galaxies.
The logarithmic scale for the S\'{e}rsic index axis makes the visual comparison with the concentration index (which is already a logarithmic quantity) easier, but at the cost of showing a bent LDA line.
The \ur color is the main discriminant, but the S\'{e}rsic index helps to separate the ET and LT sets more clearly.
The separation line is located at $\ur = (2.756 \pm 0.002) - (0.14125 \pm 0.00007) n$, where $n$ is the S\'{e}rsic index.
An example for the confusion matrix for the test set LDA classification is shown in Table~\ref{tab:ldaconf}.
For a total of \nurtest\ test galaxies, only 15 (7+8) were misclassified, yielding a classification accuracy of \FPupn\result{\nurtest{} 15 \nurtest{} - / 3 round}\(\result \pm 0.008\) in this particular case.


Linear discriminant analysis improves both the baseline and the \ur color classifications, as shown in Table~\ref{tab:compar} and Fig.~\ref{fig:dnn_acc}.
The average \sensitivity of 0.80 is similar to the \ur color, and the \specificity of 0.979 is marginally larger.
Altogether, including the S\'{e}rsic index has helped to obtain a moderate improvement on the average accuracy (from 0.96 to 0.970) but reduces the accuracy uncertainty by \FPupn\result{0.02 0.008 0.02 - / 100 * 0 round}\result \% (from 0.02 to 0.008) with respect to the \ur color discriminant.

The LDA classification presented above is a simple machine learning methodology that shows the potential of this kind of algorithm.
As with most machine learning methods, LDA does not incorporate an easy solution to deal with missing data, although the research in this area has been continuous over the last 50 years \citep[\eg][]{jac68, cha76, cai18}. 
Therefore, the usual way to deal with missing values is simply dropping incomplete records.
This is a major problem when dealing with cross-correlated data gathered from multiple catalogs because missing data is a frequent characteristic of catalog entries.
Thus, to prevent a drastic reduction in the amount of complete records, we are forced to put a limit on the number of photometric colors.


\subsubsection{\dnn classification}

Classification based on \dnns allows us to overcome the missing data problem that limits the number of feasible variables of other machine learning solutions.
This feature by itself justifies its application in astronomical databases, where records are often incomplete.
In the following, we show the results obtained for both OTELO and COSMOS photometry and S\'{e}rsic index samples.

\paragraph{OTELO\\}

We applied a very simple \dnn to the OTELO catalog.
First we computed the colors \mbox{$u-r$}, \mbox{$g-r$}, \mbox{$r-i$}, \mbox{$r-z$}, \mbox{$r-J$}, \mbox{$r-H$}, and \mbox{$r-Ks$}, and we introduced these colors as inputs in the \dnn along with the $r$ magnitude and the S\'{e}rsic index, that is, a total of nine input factors feeding the \dnn.
%
%
%
One example of the 100 random samplings analyzed with our \dnn classification is shown in Table~\ref{tab:dnn}.
For this particular example, the classification accuracy is \FPupn\result{551 27 516 + / 3 round}\(\result \pm 0.006.\)
We highlight the fact that, because of the missing data management, the number of cases included in the \dnn classification (551) is larger than those for the \ur color and LDA methods (536), despite the differences in the number of input factors (9 for the \dnn versus 2 for the LDA or 1 for the \ur color) which in most machine learning techniques would lead to  a larger number of incomplete records being left out.

The mean accuracy for our 100 \dnn samplings is \(0.985 \pm 0.007\), as shown in Table~\ref{tab:compar} and in Fig.~\ref{fig:dnn_acc}.
The \sensitivity is 0.84, marginally larger than the \ur and LDA values, and the \specificity is the highest of the three methods tested.


Table~\ref{tab:missmatch} shows the eight discrepancies between the \dnn and OTELO classifications for the test sample data set presented in Table~\ref{tab:dnn}.

\begin{figure*}[hbt]
  \begin{tikzpicture}
    \draw (0.1, 0) node[inner sep=0] {\includegraphics[width=0.47\textwidth, trim={1cm 6cm 0 9.2mm}, clip]{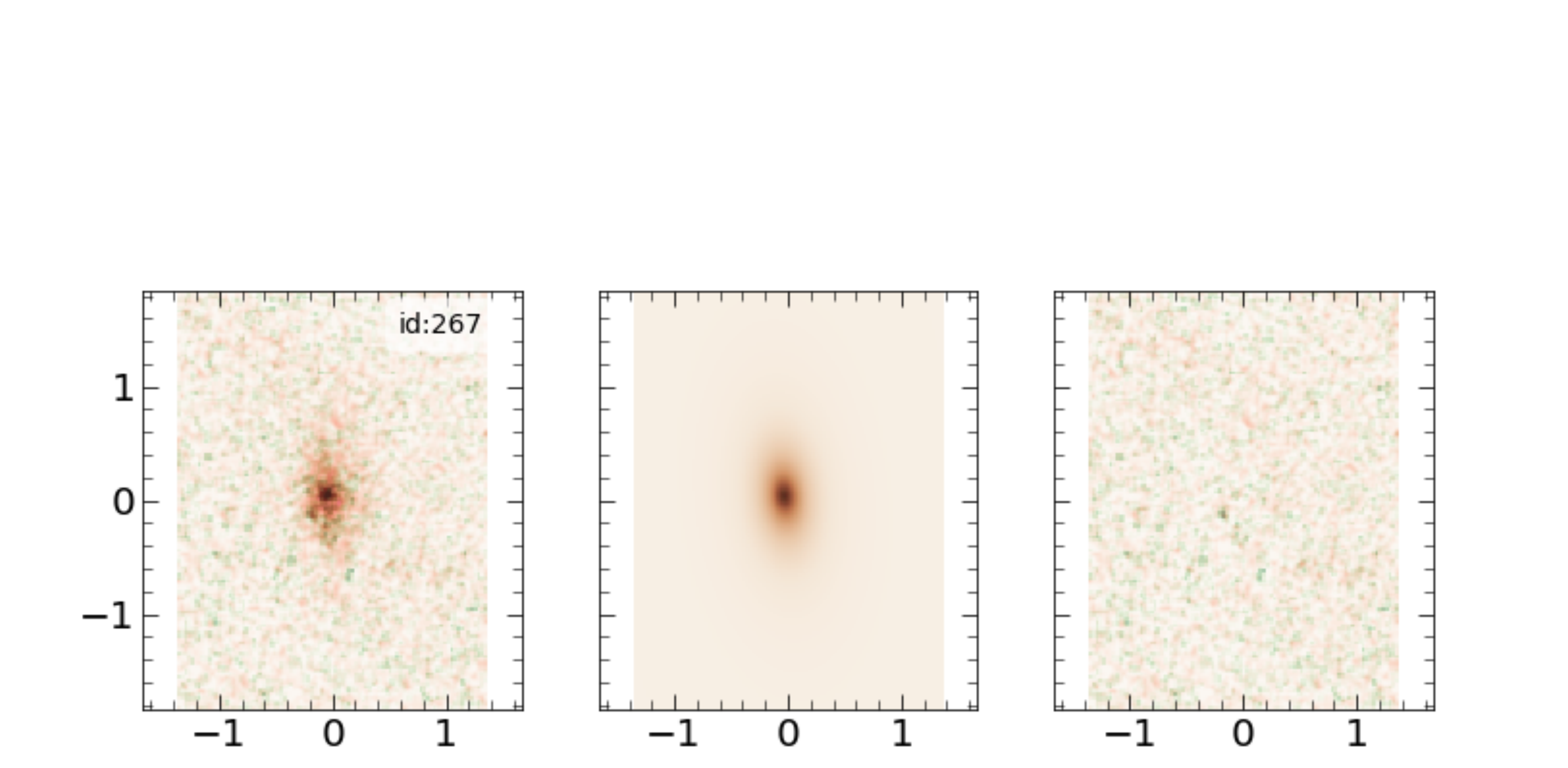}};
    \draw (9.6, 0) node[inner sep=0] {\includegraphics[width=0.47\textwidth, trim={1cm 6cm 0 9.2mm}, clip]{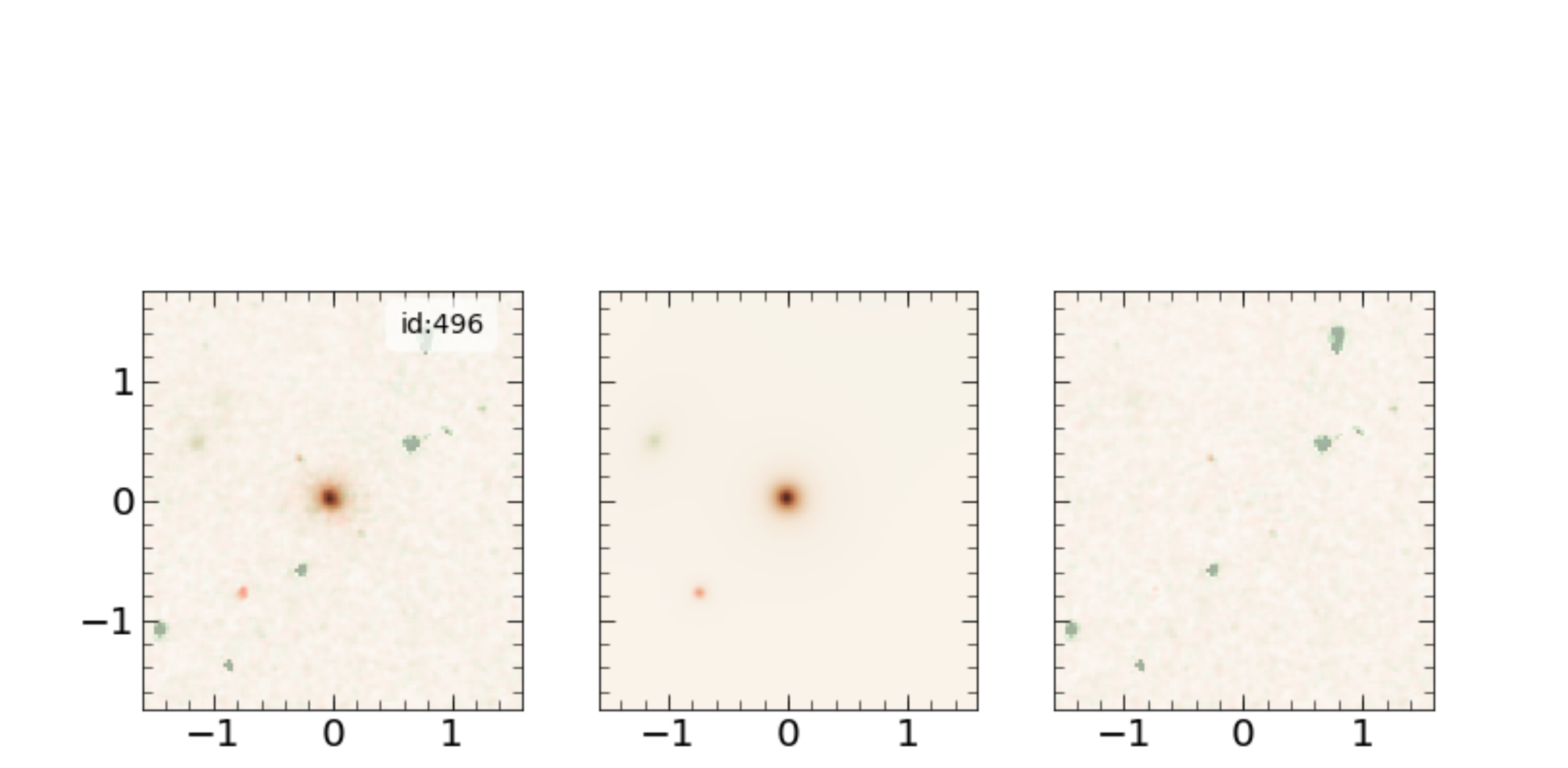}};
    
    \draw (0, -3) node[inner sep=0] {\includegraphics[width=0.48\textwidth, trim={1cm 6cm 0 9.2mm}, clip]{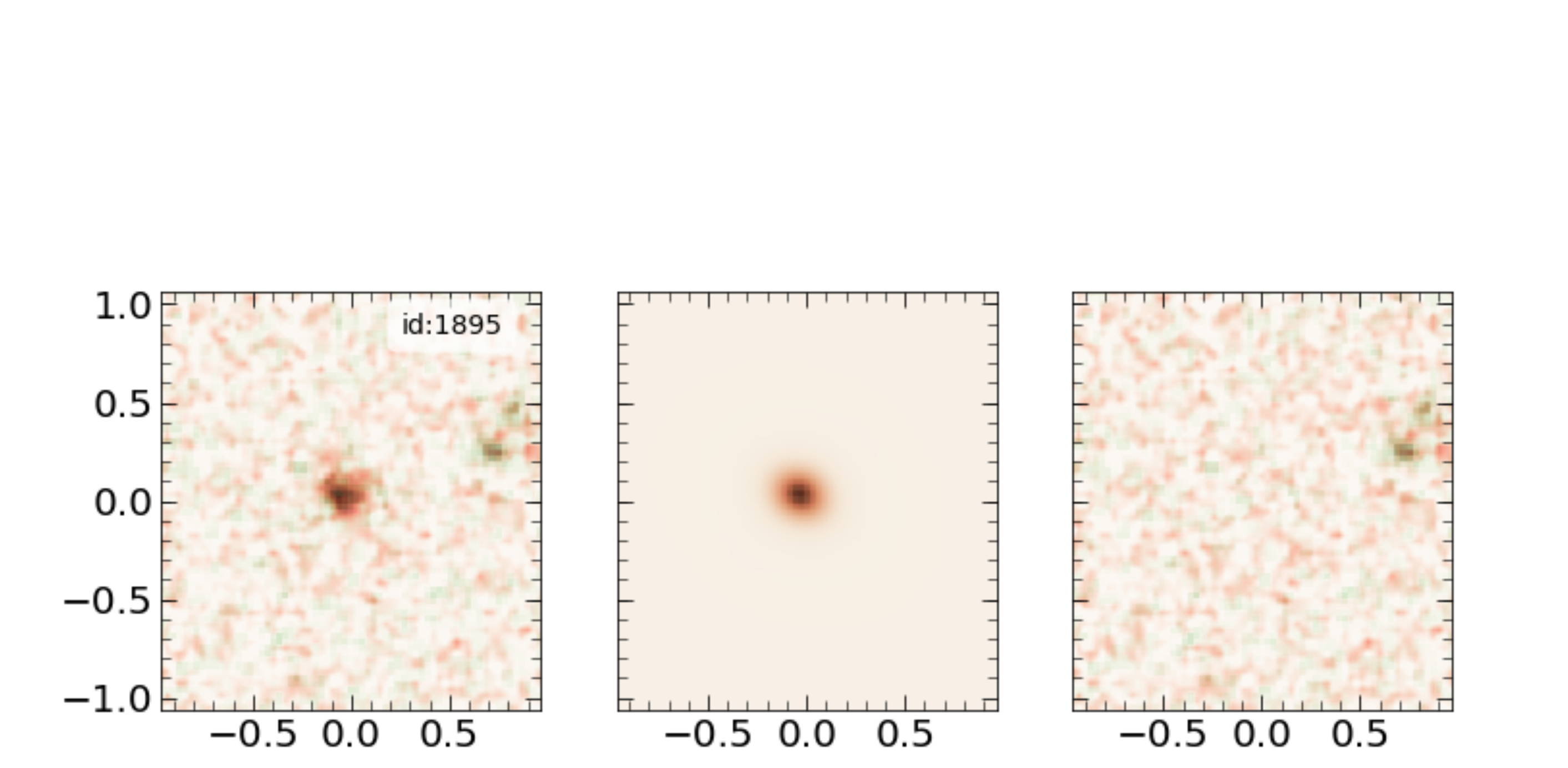}};
    \draw (9.6, -3) node[inner sep=0] {\includegraphics[width=0.47\textwidth, trim={1cm 6cm 0 9.2mm}, clip]{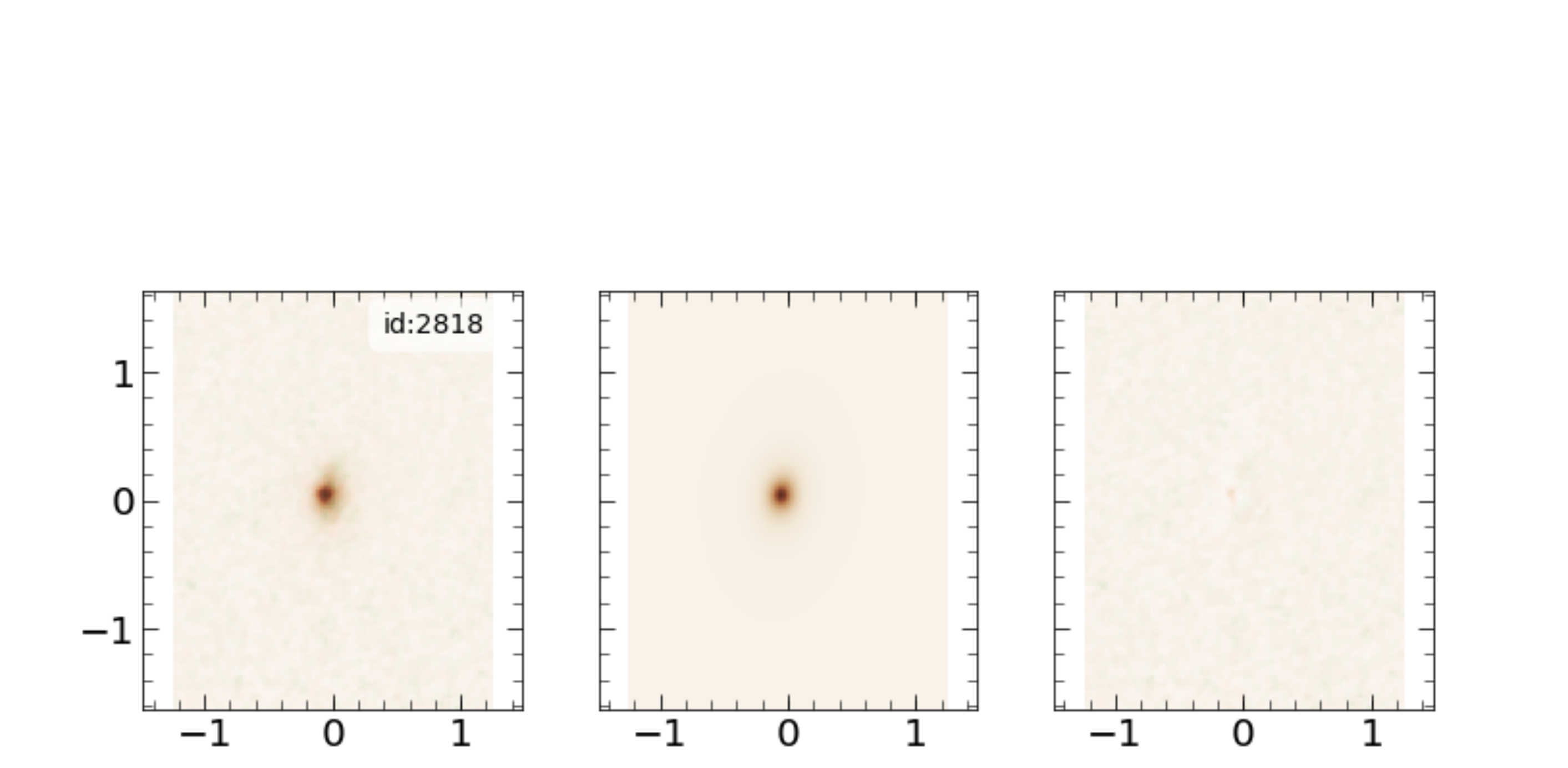}};
    
    \draw (0, -6) node[inner sep=0] {\includegraphics[width=0.48\textwidth, trim={1cm 6cm 0 9.2mm}, clip]{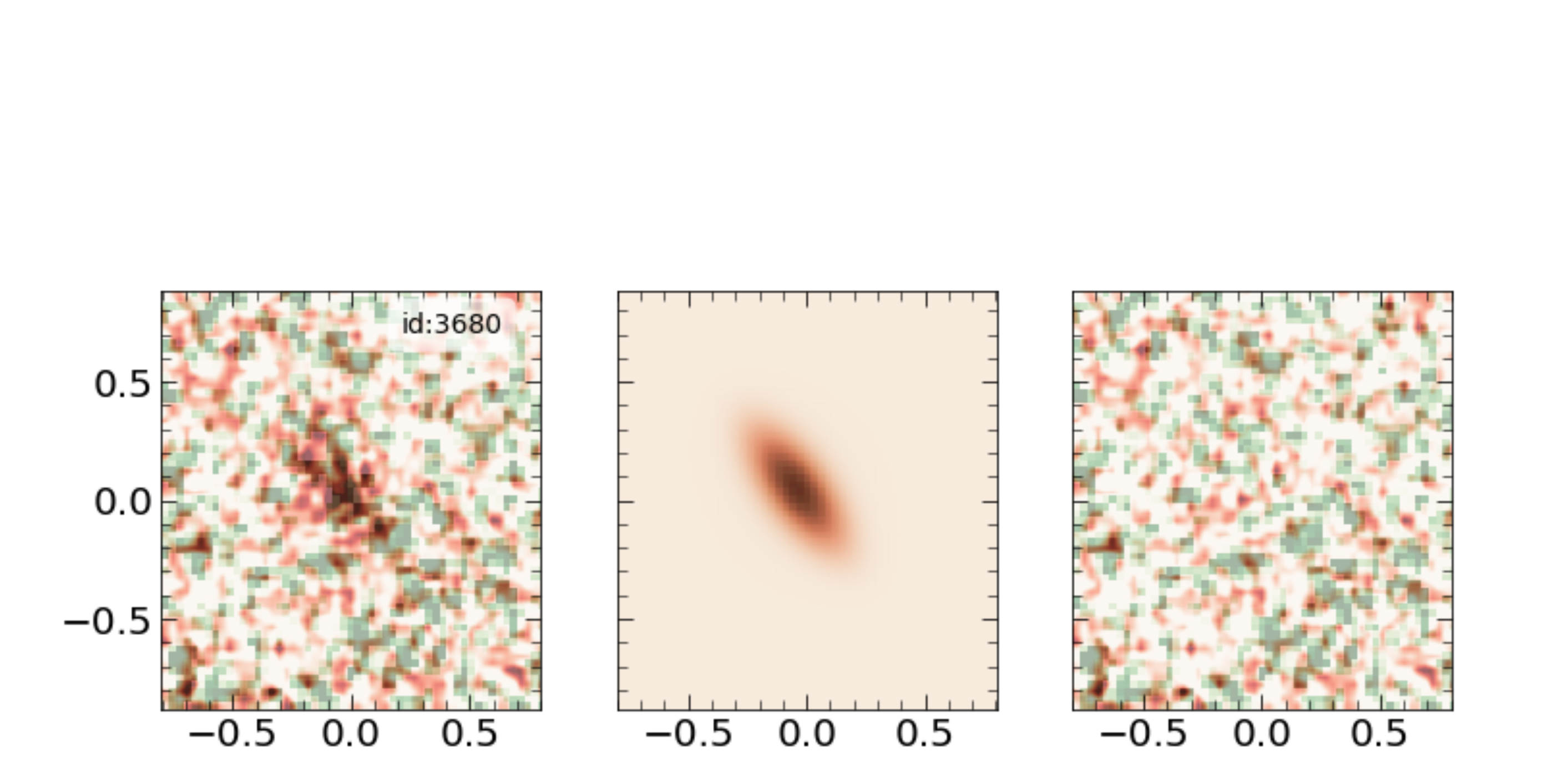}};%
    \draw (9.6, -6) node[inner sep=0] {\includegraphics[width=0.47\textwidth, trim={1cm 6cm 0 9.2mm}, clip]{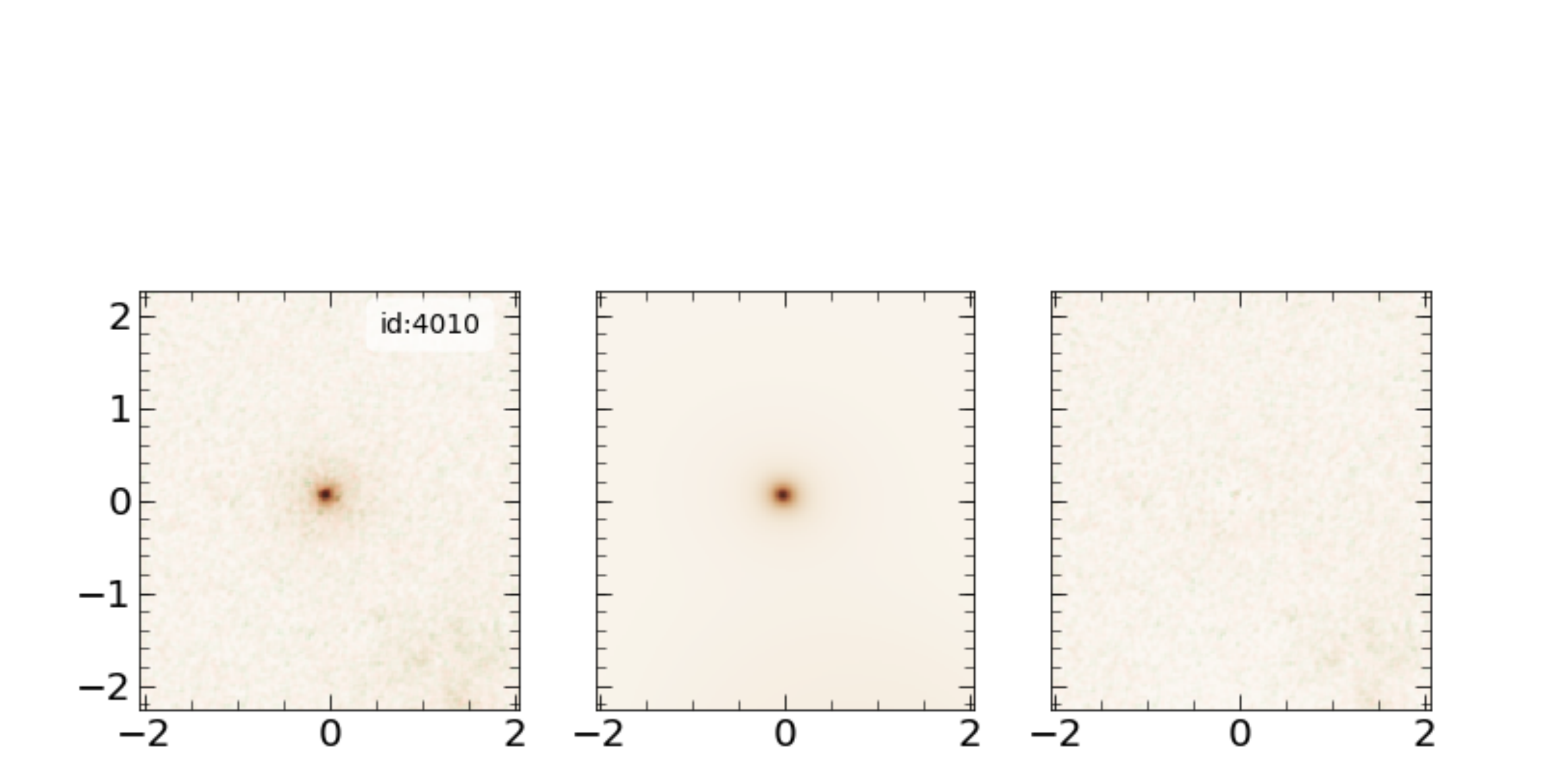}};%
    
    \draw (0.1, -9) node[inner sep=0] {\includegraphics[width=0.47\textwidth, trim={1cm 6cm 0 9.2mm}, clip]{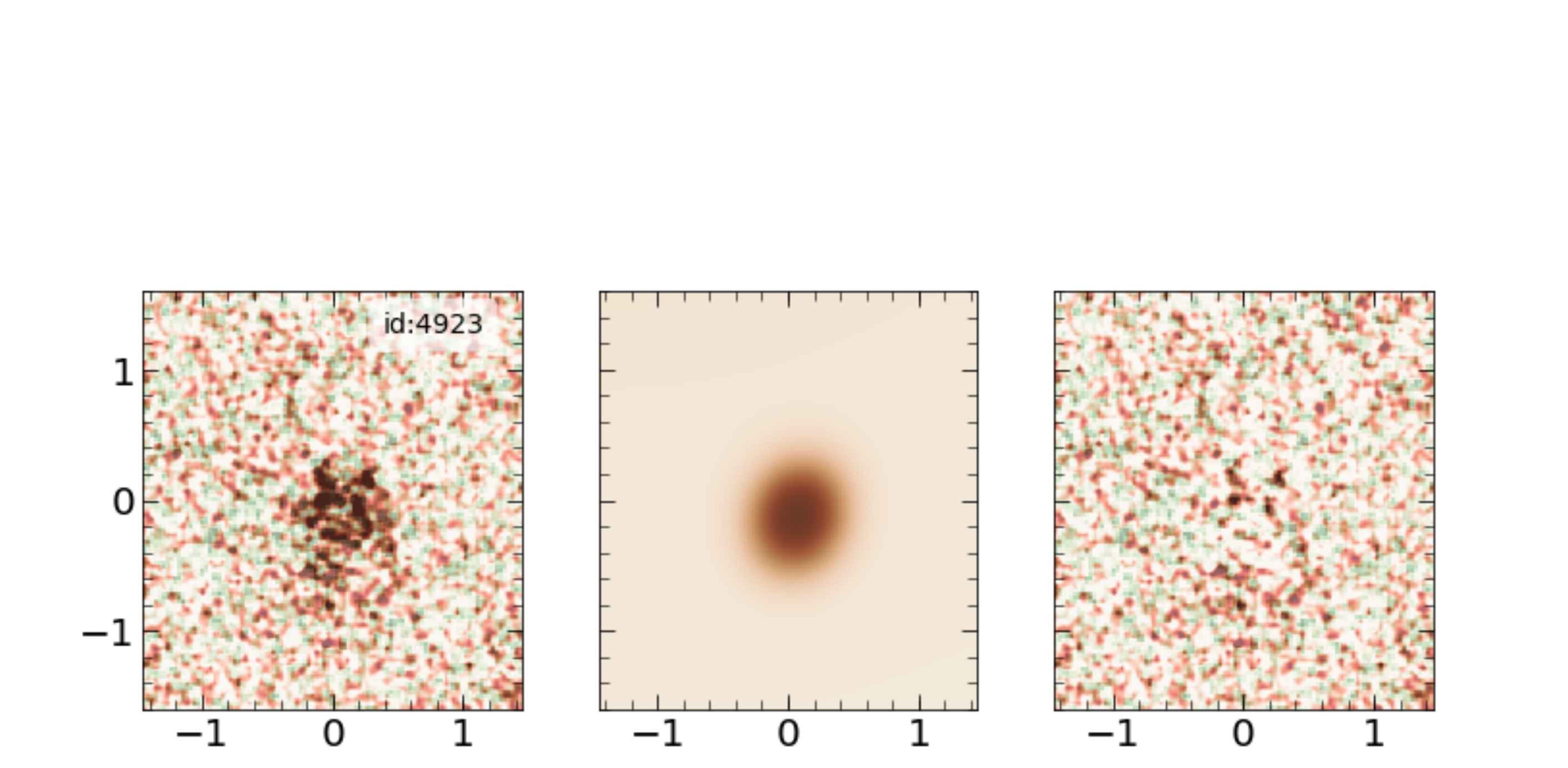}};%
    \draw (9.5, -9) node[inner sep=0] {\includegraphics[width=0.48\textwidth, trim={1cm 6cm 0 9.2mm}, clip]{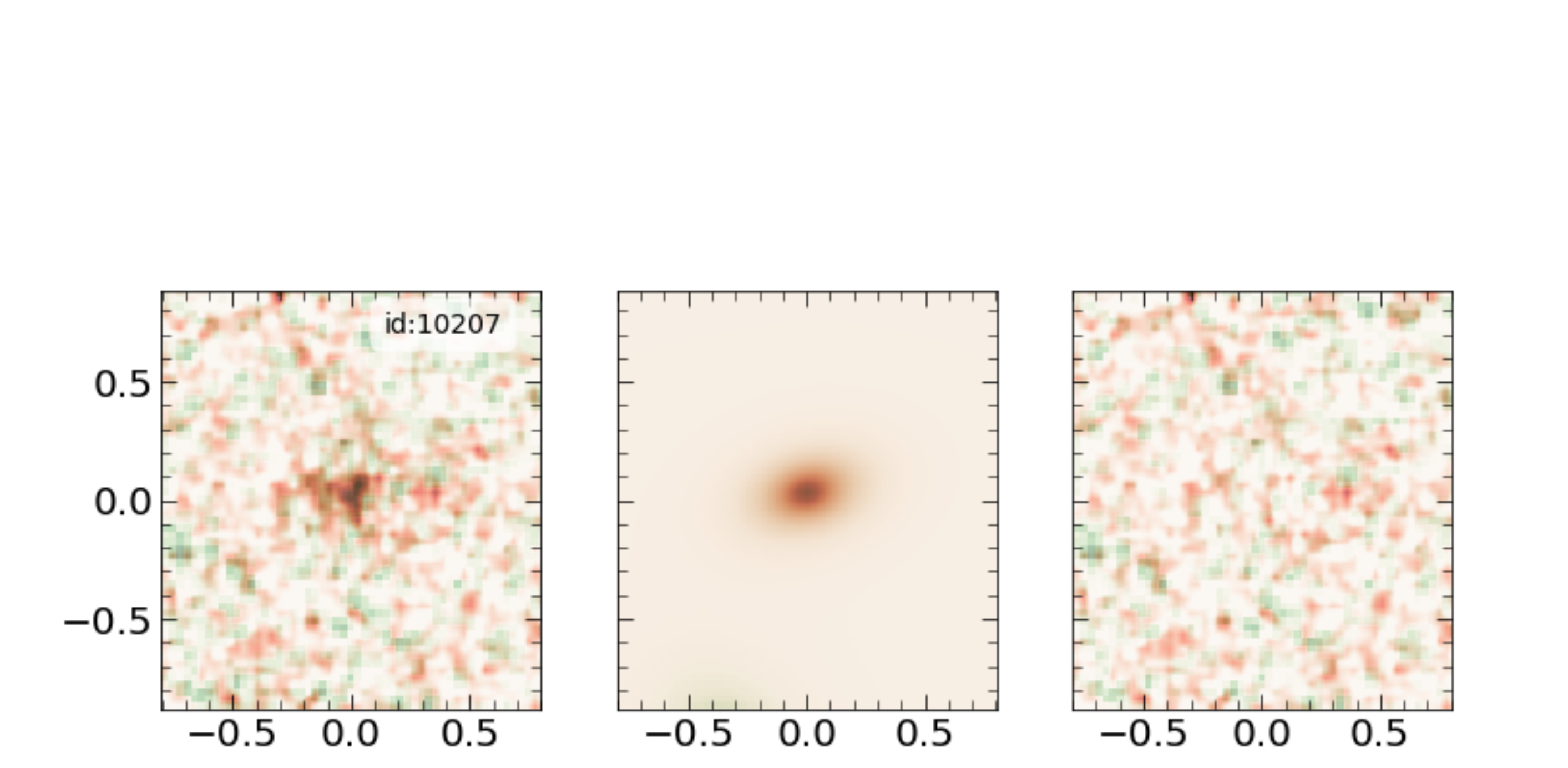}};%

  \end{tikzpicture}
  \caption{OTELO -- DNN discrepancies. Relative declination vs. right ascension coordinates in arcsec. \emph{First column:} Combined HST images from the F814W (reddish) and F606W (greenish) bands, with galaxy ID at the top-right corner. \emph{Second column:} GALFIT models for the light distribution. \emph{Third column:} HST minus GALFIT model residuals. \emph{Columns fourth, fifth and sixth} repeat the order of the previous columns.}\label{fig:galaxies}
\end{figure*}

Figure~\ref{fig:galaxies} presents the HST images in the F814W band for these eight galaxies.
For the visual classification, we have taken into account the galaxy elongation and the light distribution in the HST image; the GALFIT model helps to indicate the shape and orientation, and the image residuals indicate a possible lack of fitting or possible substructures not visible in the HST image.
Elongated and fuzzy images support a LT visual classification, while a round and soft appearance points to an ET galaxy.
From our visual check, we conclude that six out of the eight galaxies with different class ascription are correctly classified by our \dnn algorithm.
Following is a brief description of each mismatched object.

\begin{figure*}[hbt]
  \begin{tikzpicture}
    \draw (0, 0) node[inner sep=0] {\includegraphics[width=0.48\textwidth, trim={1cm 6cm 0 9.2mm}, clip]{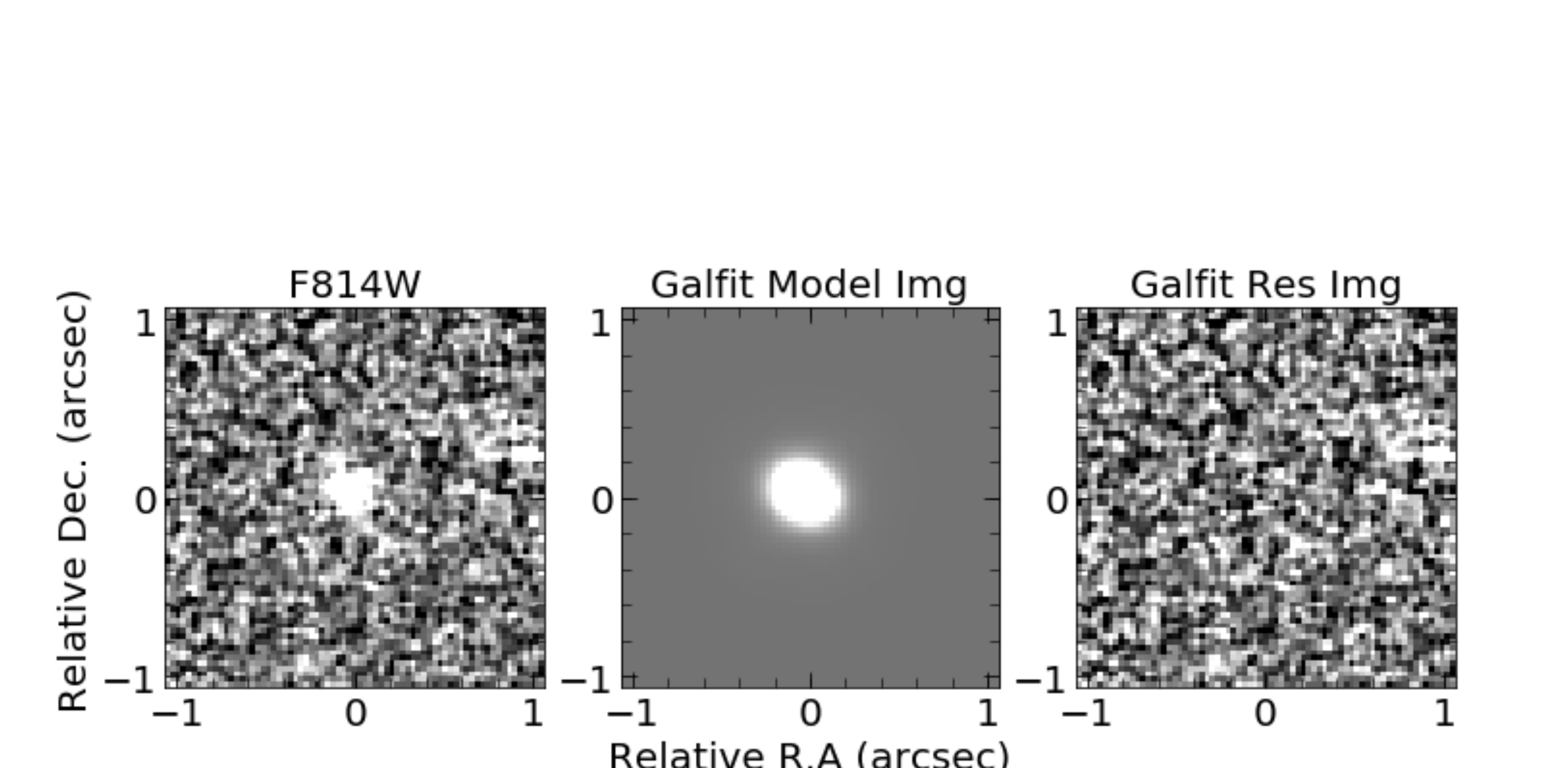}};
    \draw (9.5, 0) node[inner sep=0] {\includegraphics[width=0.48\textwidth, trim={1cm 6cm 0 9.2mm}, clip]{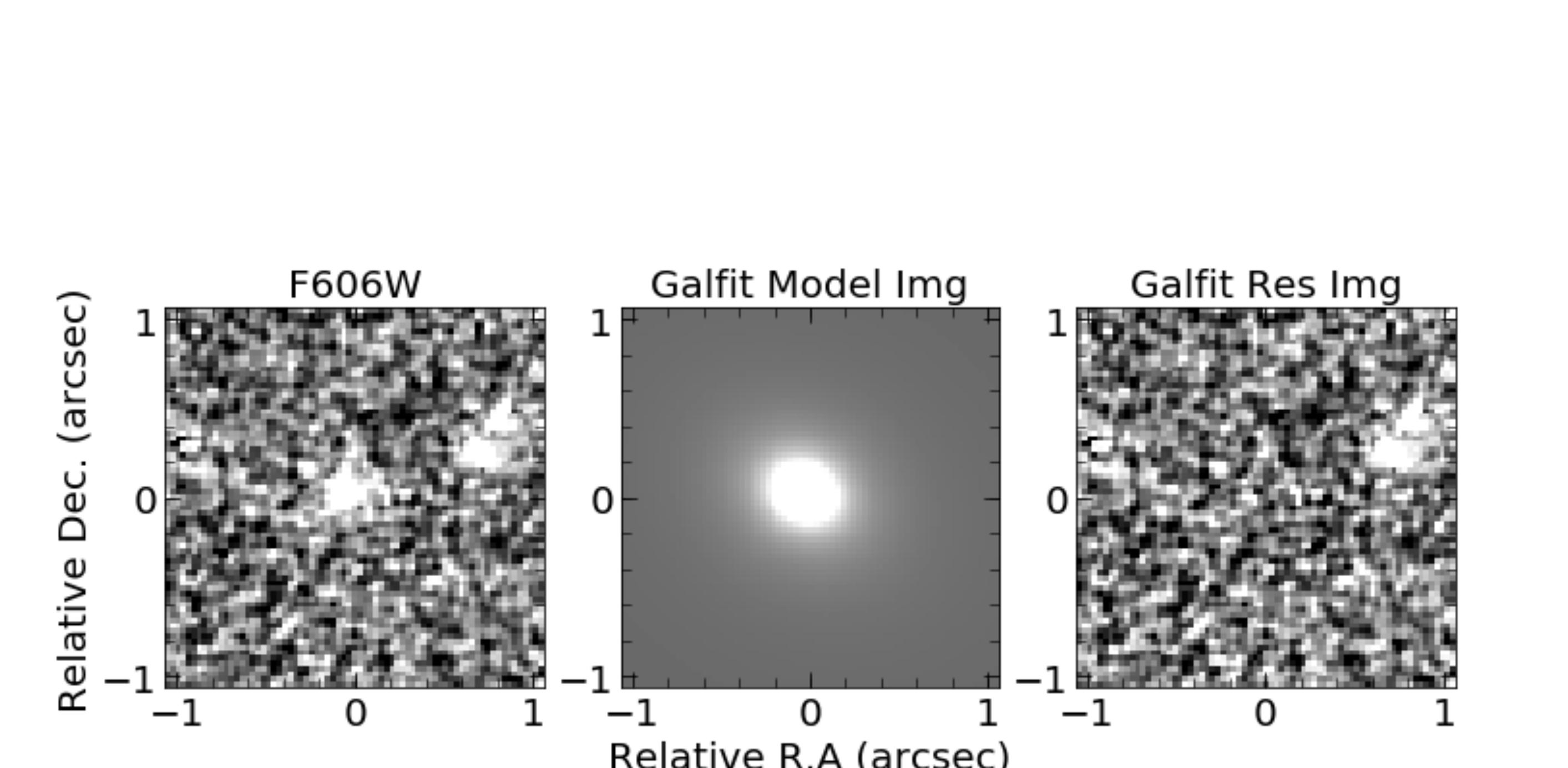}};
  \end{tikzpicture}
  \caption{HST images of ID 1895 and GALFIT models. \emph{Fist column:} High-resolution HST images in the F814W band. \emph{Second column:} GALFIT models for the light distribution. \emph{Third column:} HST minus GALFIT model residuals. \emph{Columns fourth, fifth and sixth} as the previous columns for the HST F606W images. There is a fuzzy object near the eastern border of the F606W image that is not resolved in the OTELO deep image.}\label{fig:id1895}
\end{figure*}

\begin{itemize}
  \item {\bf ID 267.} A north--south oriented disk galaxy with a fuzzy northeast portion. 
		The bulge of the galaxy broadly dominates the disk component. 
		Compatible with a Sab class.
        Visual classification as LT.
  \item {\bf ID 496.} A rounded smooth galaxy with a visual LT companion at the northwest and a star at the southwest.
        Visual classification as ET.
  \item {\bf ID 1895.} Appears as a rounded and compact galaxy in the HST F814W image (our detection image used for GALAPAGOS). 
		However, visual inspection of the HST F606W image shown in Fig.~\ref{fig:id1895} reveals that there is a companion source not detected in F814W. 
		Using our web-based graphic user interface\footnote{\url{http://research.iac.es/proyecto/otelo/pages/data-tools/analysis.php}} we find that this companion, probably a LT galaxy neither detected in the OTELO deep image, enters the ellipse which was used to extract photometry. 
		It is likely that a composite SED could be well fitted by LT templates instead single-population one.
        Visual classification as ET with an unresolved companion.
  \item {\bf ID 2818.} A round shaped galaxy, the image of residuals suggests possible over-subtraction.
        Visual classification as ET.
  \item {\bf ID 3680.} A small, fuzzy, northwest to southeast oriented disk galaxy.
        Visual classification as LT
  \item {\bf ID 4010.} A rounded galaxy, with a LT companion at the southeast.
        Visual classification as ET.
  \item {\bf ID 4923.} A faint and fuzzy galaxy.
        Visual classification as LT.
  \item {\bf ID 10207.} A west-east oriented fuzzy small disk galaxy.
        Visual classification as LT.
\end{itemize}




\paragraph{COSMOS\\}

We used the COSMOS dataset to check for the reliability of our \dnn architecture.
Using the ZSMC and CPhR catalogs we built a sample of \ngalC galaxies for which the photometry and S\'{e}rsic indexes are available.
Photometric bands in the CPhR catalog do not exactly match OTELO's bands.
We have included Subaru's $BV$ bands, and have excluded the $H$ band which is absent from the CPhR database.
Thus, the COSMOS data used in this work consist of nine photometric bands (compared with eight in the case of OTELO catalog) and the S\'{e}rsic index.
Because the OTELO and COSMOS bands are different, we had to train our \dnn model again.
As in the case of OTELO, we fed the \dnn with the S\'{e}rsic index, the $r$ magnitudes, and the colors relative to the $r$ band.
We did so without changing the \dnn architecture except for the number of inputs.
Despite the differences between the two datasets, we shall see that our \dnn architecture reaches a high classification accuracy also for the COSMOS data.

\begin{table}[t]
\caption{COSMOS \dnn confusion matrix}\label{tab:cosmos}
\centering
\begin{tabular}{llrrr}
  \hline
  \noalign{\smallskip}
 &        & \multicolumn{2}{c}{COSMOS} \\
  \cline{3-4}
  \noalign{\smallskip}
 & & ET & LT & \textbf{Total} \\ 
  \noalign{\smallskip}
  \hline
  \noalign{\smallskip}
  \multirow{2}{*}{\dnn} & ET & 4450 &   457  & \textbf{ 4907} \\ 
                        & LT &  517 & 24264  & \textbf{24781} \\ 
                       & \textbf{Total} 
                                & \textbf{4967} & \textbf{24721} & \textbf{29688} \\ 
   \noalign{\smallskip}
   \hline
\end{tabular}
\end{table}




\begin{table*}[ht]
\caption{Comparison of classification methods for \pc samples.}\label{tab:compar_concent}
\centering
\begin{tabular}{llllllllllr}
  \hline
  \noalign{\smallskip}
  \multirow{2}{*}{Database} & \multirow{2}{*}{Method} & \multicolumn{2}{c}{Accuracy$\,^a$} & & \multicolumn{2}{c}{\sensitivity$\!^a$} & & \multicolumn{2}{c}{\specificity$\!^a$} & \multicolumn{1}{c}{Sample} \\
  \cline{3-4}
  \cline{6-7}
  \cline{9-10}
  \noalign{\smallskip}
 && Mean & Error & & Mean & Error & & Mean & Error & \multicolumn{1}{c}{size} \\ 
  \noalign{\smallskip}
  \hline
  \noalign{\smallskip}
  OTELO & Baseline  & \FPupn\result{\pltc{} 100 swap / 3 round}\result & 0.005 && 0 & ~\dots && 1 & ~\dots &  \ngalc \\ 
  OTELO & \ur color & 0.96 & 0.01 && 0.7 & 0.4 && 0.98 & 0.02 & 657 \\ 
  OTELO & LDA       & 0.971 & 0.007 && 0.78 & 0.09 && 0.980 &  0.006 & 657 \\
  OTELO & \dnn       & 0.980 & 0.006 && 0.75 & 0.09 && 0.992 & 0.005 & 688 \\
  \noalign{\smallskip}
  COSMOS & Baseline  & \FPupn\result{\pltCc{} 100 swap / 3 round}\result & 0.001 && 0 & ~\dots && 1 & ~\dots &  \ngalCc \\ 
  COSMOS & \dnn       & 0.971 & 0.001 && 0.84 & 0.03 && 0.985 & 0.004 & \ntestCc \\ 
  \noalign{\smallskip}
  \hline
  \noalign{\smallskip}
  \multicolumn{4}{l}{\footnotesize{$^a$ On 100 bootstrap runs.}}
\end{tabular}
\end{table*}

\begin{figure*}
   \centering
   \includegraphics[width=0.5\textwidth]{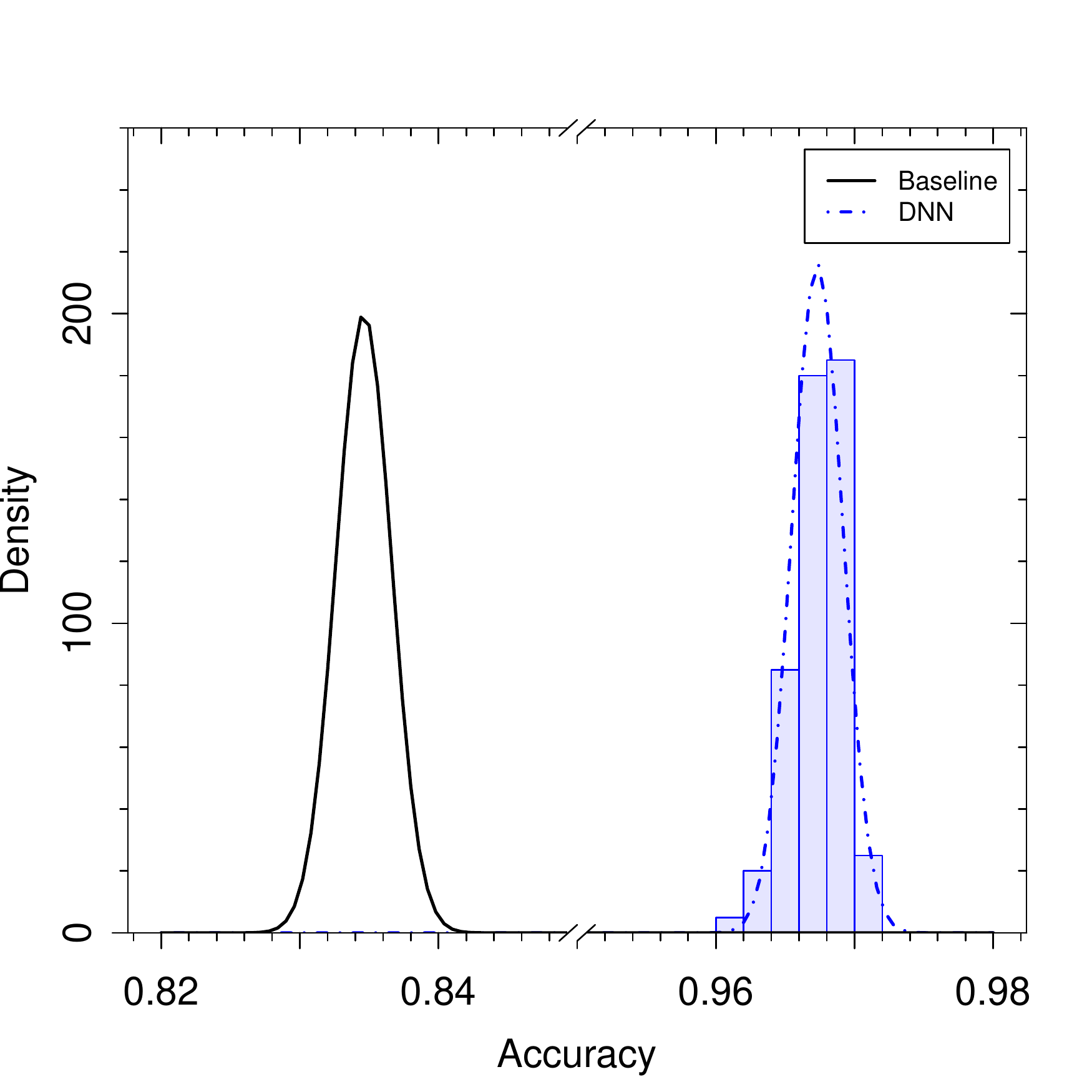}
   \caption{Histogram of accuracies for 100 galaxy classification bootstrap runs using the COSMOS \ps sample. 
   The solid black line corresponds to the baseline accuracy distribution of a sample of \ngalC COSMOS galaxies.
   The blue histogram and dash-dotted line shows the \dnn accuracy distribution for a test sample of \ntestC galaxies.}\label{fig:dnn_cosmos}%
\end{figure*}

Table \ref{tab:cosmos} shows the confusion matrix for one of the 100 random samplings that we used to characterize the COSMOS \dnn.
For this sampling in particular, the classification accuracy is of \FPupn\result{\soltestC{} 4450 24264 + / 3 round}\FPupn\err{\result{} 1 - \result{} * \soltestC{} swap / 2 swap root 3 round}$\result \pm \err.$
Figure \ref{fig:dnn_cosmos} and Table \ref{tab:compar} show the distribution of accuracies for 100 \dnn classification trials obtained from the COSMOS dataset.
The mean accuracy for these trials is $0.967 \pm 0.002$, well above the relatively low baseline of \FPupn\result{\solnC{} \nltC{} / 3 round}\FPupn\err{\result{} 1 - \result{} * \solnC{} swap / 2 swap root 3 round}$\result \pm \err,$ which corresponds to \numprint{28951} LT galaxies out of a total of \ngalC objects included in our COSMOS \ps sample.
Not only is the COSMOS \ps baseline lower than OTELO's (0.946), but the \dnn performance is also lower: 0.967 for COSMOS compared with 0.985 for OTELO.
The \sensitivity of 0.91 for COSMOS \ps is similar within the errors to that of OTELO (0.84), but the \specificity for COSMOS is slightly lower (0.979) than that for OTELO (0.993).

Applying the same \dnn architecture to the OTELO and COSMOS datasets, the method yields high classification accuracy in both cases.
Band differences between both datasets may contribute to the accuracy results.
We note that OTELO optical bands were gathered from CFHTLS data, but most of the COSMOS optical bands used were measured by Subaru.
The OTELO $H$ band is missed in COSMOS, while the COSMOS $BV$ bands, which are not included in our OTELO dataset, are heavily correlated to $gr$ bands. 

The high classification accuracies for both the OTELO and the COSMOS datasets suggests that our proposed \dnn architecture may be applicable to a large number of databases that encompass both visual and infrared photometric bands and an estimate of the S\'{e}rsic index.

\subsection{\cp samples}

        The S\'{e}rsic index that we used in the LDA and \dnn classification methods detailed above is obtained through a parametric fitting that is difficult to achieve when dealing with low-resolution images.
        On the contrary, the radius containing a given fraction of the galaxy total brightness is easier to estimate and can be measured directly.
        In this section we repeat our previous analysis of the OTELO and COSMOS databases, but using samples obtained through the concentration index defined as the ratio between the radii containing 80\% and 20\% of the galaxy brightness.
        
        Table~\ref{tab:compar_concent} shows the results obtained with the OTELO and COSMOS CP samples. 
        As in the S\'{e}rsic index samples, the \dnn classification yields the highest accuracy for OTELO (0.980), and  also yields very accurate results for COSMOS (0.971).
        In general, the results are comparable with those obtained using the SP sample.
        
\begin{figure*}
   \centering
   \includegraphics[width=0.5\textwidth]{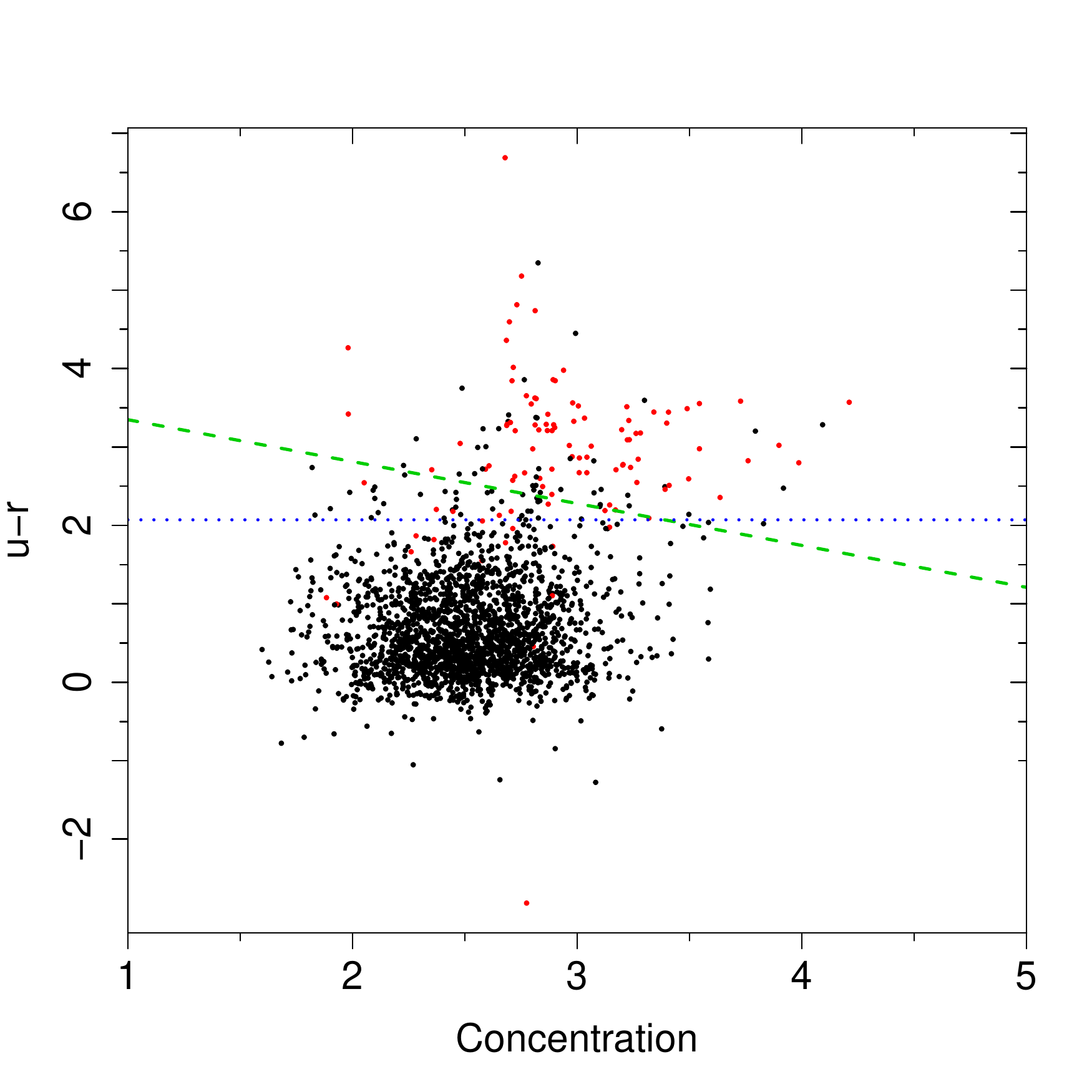}
   \caption{Classification for the OTELO \pc sample of \ngalc galaxies through the \ur color and LDA algorithms. The \ur color vs. concentration plot shows the original morphological type classification in the OTELO catalog reduced to LT galaxies (black circles) and ET galaxies (red squares). The dotted blue line indicates the \ur color separation, and the dashed green line the LDA separation by the \ur color and the concentration. The reader should take into account that this is not a flux limited sample.}\label{fig:lda_conc}%
\end{figure*}

\begin{figure*}
   \centering
   \includegraphics[width=0.5\textwidth]{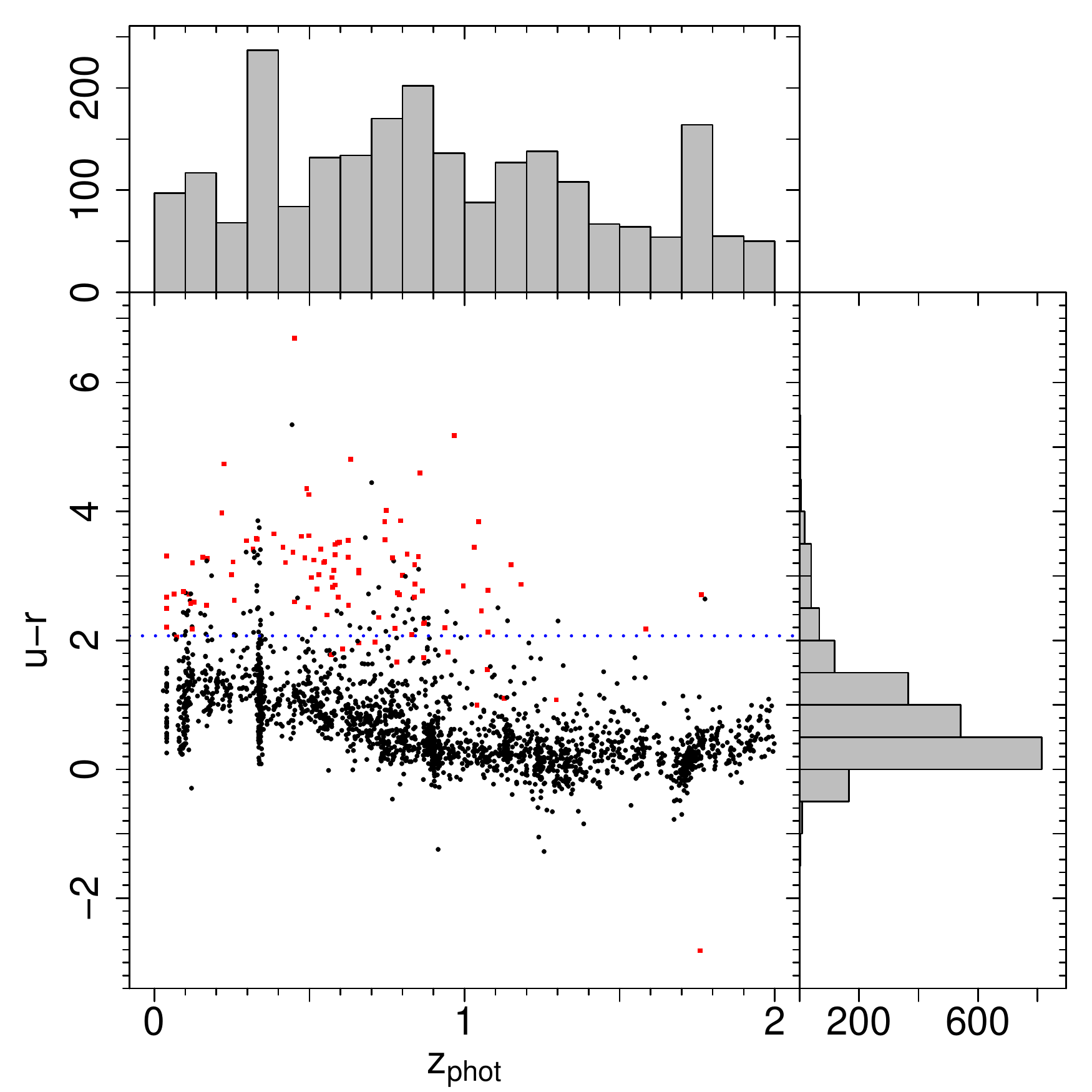}
   \caption{As Fig. \ref{fig:ur_z_sp} but for the \ngalc galaxies in the OTELO \pc sample.} \label{fig:ur_z_cp}%
\end{figure*}

        Figure~\ref{fig:lda_conc} shows the distribution of the \ur colors versus the concentration index, along with the \ur color and the LDA separation boundaries.
        The \ur color separation is $2.1 \pm 0.3$, in agreement with  the values for the OTELO \ps sample ($2.0 \pm 0.2)$, and \citet[$\ur = \urplane$]{str01}. 
        The LDA separation is located at 
        $$u-r = (3.882 \pm 0.002) - (0.5342 \pm 0.0003) C,$$ 
        where $C$ is the concentration.
        The same trend of LT galaxies getting bluer at redshifts $z > 0.5$ can be seen in Fig. \ref{fig:ur_z_cp}.
                
\begin{figure*} 
  \centering
  \includegraphics[width=0.5\textwidth, clip]{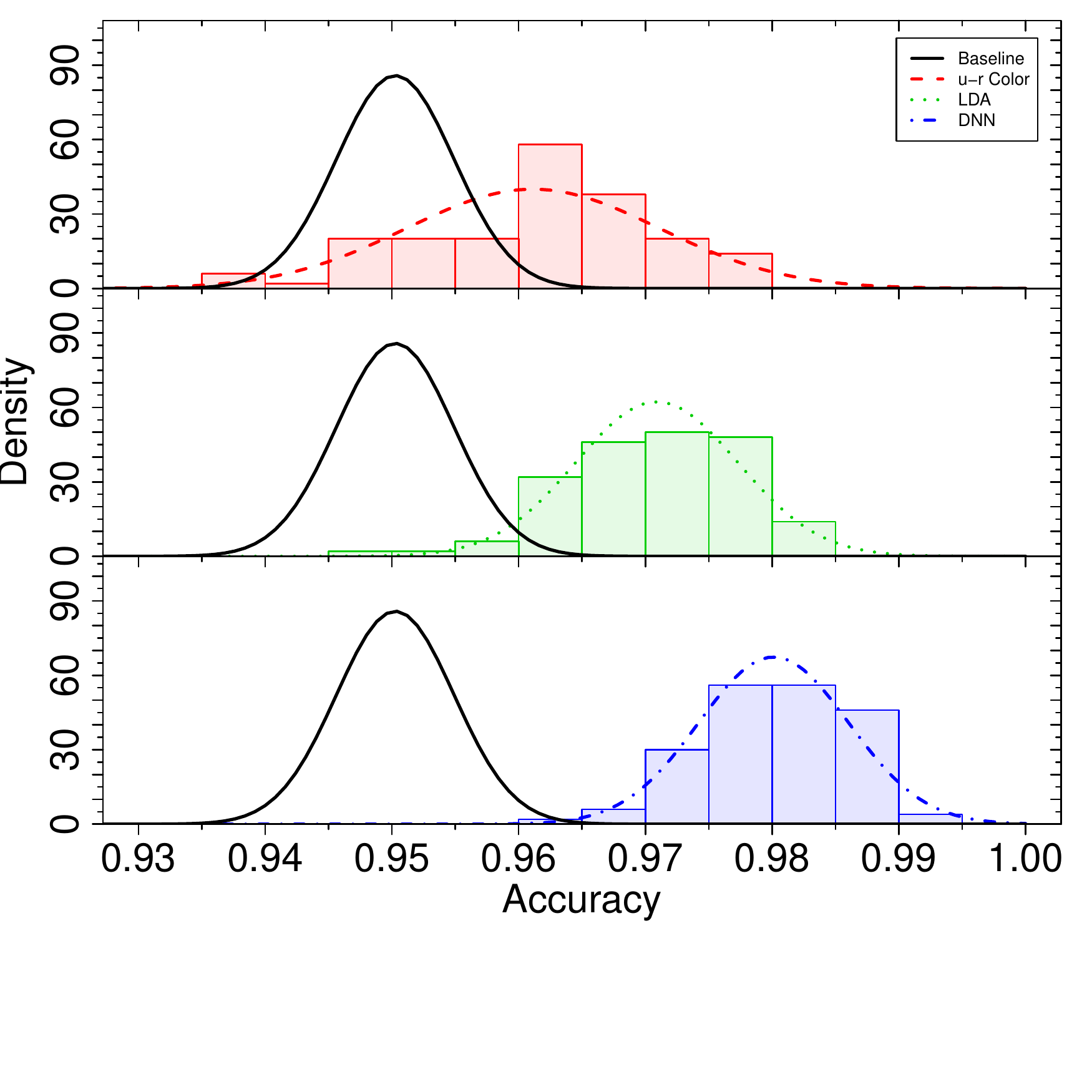}
  \caption{As Fig. \ref{fig:dnn_acc} but for the 100 galaxy classification bootstrap runs for the \ngalc galaxies of the OTELO \pc sample.}\label{fig:conc_dnn_acc}%
\end{figure*}

        Figure~\ref{fig:conc_dnn_acc} shows the distributions for the accuracies of the baseline, \ur color, LDA and \dnn classifications performed on the OTELO CP sample.
        As in the OTELO SP sample, the \dnn yields the best accuracy, then LDA and finally the \ur color classification.
        
\begin{figure*} 
  \centering
  \includegraphics[width=0.5\textwidth]{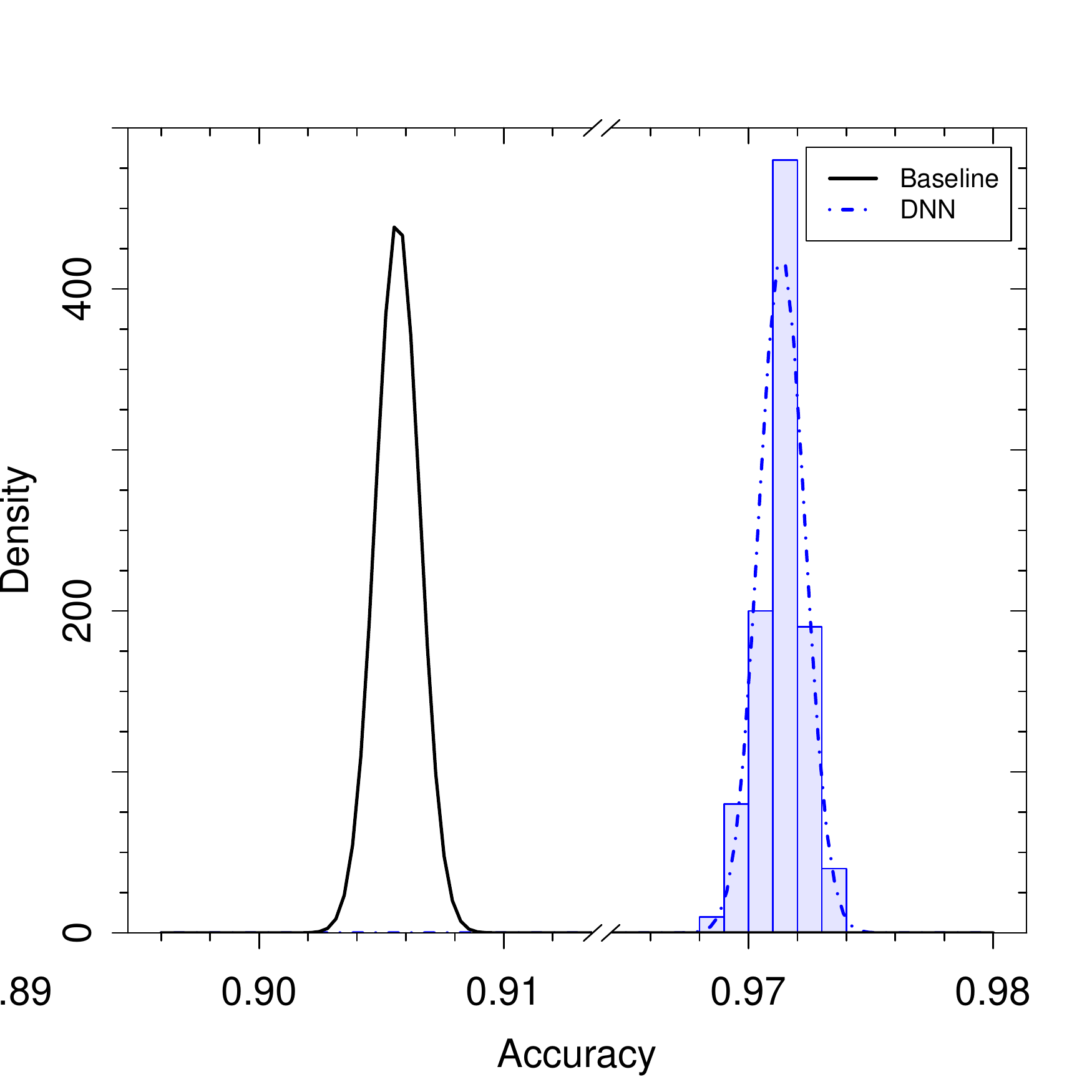}
  \caption{Histogram of accuracies for 100 galaxy classification bootstrap runs using the COSMOS \pc sample..
   The solid black line corresponds to the baseline accuracy distribution of a sample of \ngalCc COSMOS galaxies.
   The blue histogram and dash-dotted line shows the \dnn accuracy distribution for a test sample of \ntestCc galaxies.}\label{fig:conc_dnn_cosmos}
\end{figure*}

        The distribution of \dnn accuracies for the COSMOS \pc sample is shown in Fig.~\ref{fig:conc_dnn_cosmos}.
        Compared with the COSMOS \ps sample, the proportion of LT galaxies is larger (\numprint{\nltCc} LT out of \ngalCc galaxies), yielding a more accurate baseline (\FPeval\result{round(\nltCc{} / \solnCc{}, 3)}\result).
        The \dnn accuracies are comparable, with a lower \sensitivity and a marginally larger \specificity for the COSMOS \pc sample.
        


\subsection{Magnitude limited samples}

\begin{figure*} 
  \centering
  \includegraphics[width=0.5\textwidth]{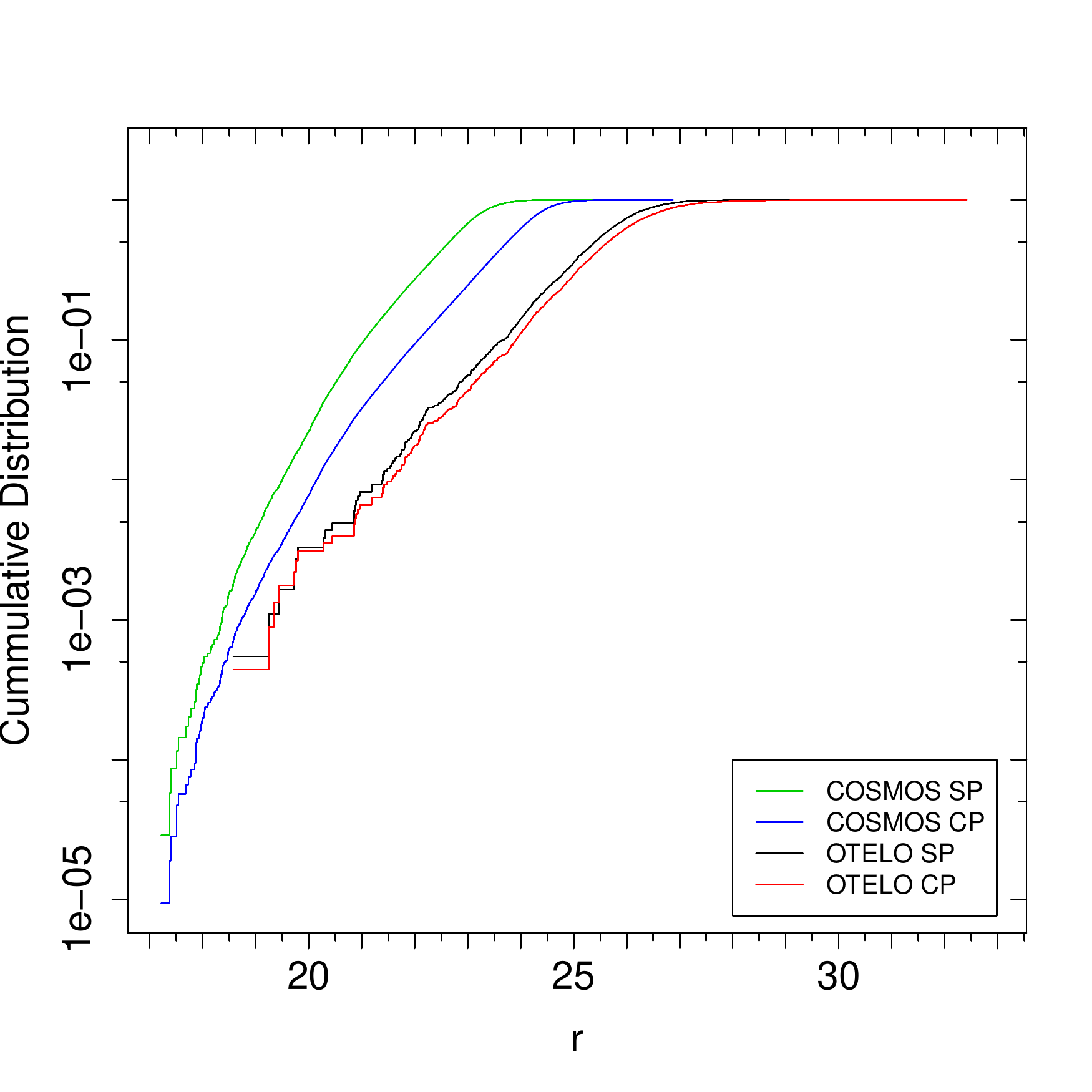}
  \caption{Cumulative distribution of galaxies by $r$ magnitudes. OTELO \ps and \pc samples are 3 and 2 magnitudes deeper than the COSMOS \ps and \pc samples, respectively. However, COSMOS sweeps a larger volume as the brightest end of the cumulative distribution implies.}\label{fig:cum_dist}
\end{figure*}

Our aim in this paper is to use machine learning techniques to distinguish between ET and LT galaxies.
Thus, our samples are selected from a redshift limited region with the only requirement of containing enough galaxies in every redshift interval for accurate training and testing the machine learning algorithm.
However, neither OTELO nor COSMOS \ps and \pc samples were flux limited to produce a complete sample of galaxies in the volume defined by $z \leq 2$.
This leads us to question the possible cosmological inferences of our results.
In this section, we present the results of the machine learning algorithms but using flux limited samples for both training and testing sets.      
Figure~\ref{fig:cum_dist} shows the cumulative distribution of galaxies by $r$ magnitudes for all the samples analyzed so far.
With respect to the low-brightness tails, both OTELO \ps and \pc samples have similar cumulative distributions that flatten around a magnitude of $r \simeq 26$.
This flattening may be considered as a rough measurement of completeness.
Thus, compared with OTELO-Deep image measurements, \citet{bon19} estimate that the OTELO catalog reaches a 50\% completeness flux at magnitude 26.38.
For COSMOS, the \ps sample flattens around $r \simeq 23$, while the \pc sample does at $r \simeq 24$.
Since COSMOS samples cover a large sky volume, their high brightness tails extend to galaxies approximately 1.5 magnitudes brighter than the much more confined OTELO volume.

\begin{table*}[ht]
\caption{Mean accuracy for OTELO samples at different magnitude limits in the r-band} 
\label{tab:otelo_samples}
{\centering
\begin{tabular}{lrrr@{$\,\pm\,$}lr@{$\,\pm\,$}lr@{$\,\pm\,$}lr@{$\,\pm\,$}lr}
  \hline
  \noalign{\smallskip}
  $r_{lim}$ & DC & $N$  & \multicolumn{2}{c}{ Baseline } & \multicolumn{2}{c}{ $u-r$ } & \multicolumn{2}{c}{ LDA } & \multicolumn{2}{c}{ DNN } \\
  \noalign{\smallskip}
 \hline
  \noalign{\smallskip}
  \multicolumn{11}{c}{OTELO SP sample} \\
  \noalign{\smallskip}
  29.36 & 0 & 1834 & 0.946 & 0.006 & 0.96 & 0.02 & 0.970 & 0.008 & 0.985 & 0.007 \\ 
  27.00 & 0.25 & 1765 & 0.947 & 0.006 & 0.96 & 0.02 & 0.969 & 0.009 & 0.977 & 0.008 \\ 
  26.00 & 0.65 & 1358 & 0.943 & 0.007 & 0.96 & 0.02 & 0.97\phantom{0} & 0.01 & 0.978 & 0.009 \\ 
  25.00 & 0.94 & 654 & 0.91\phantom{0} & 0.02 & 0.96 & 0.03 & 0.96\phantom{0} & 0.02 & 0.96\phantom{0} & 0.02 \\ 
  \noalign{\smallskip}
  \multicolumn{11}{c}{OTELO CP sample} \\
  \noalign{\smallskip}
  32.42 & 0 & 2292 & 0.950 & 0.006 & 0.96 & 0.02 & 0.971 & 0.007 & 0.980 & 0.007 \\ 
  27.00 & 0.25 & 2051 & 0.951 & 0.006 & 0.96 & 0.02 & 0.972 & 0.008 & 0.981 & 0.007 \\ 
  26.00 & 0.65 & 1437 & 0.944 & 0.007 & 0.96 & 0.02 & 0.973 & 0.008 & 0.977 & 0.009 \\ 
  25.00 & 0.94 & 661 & 0.91\phantom{0} & 0.02 & 0.96 & 0.03 & 0.96\phantom{0} & 0.02 & 0.97\phantom{0} & 0.02 \\ 
  \noalign{\smallskip}
  \hline
  \noalign{\smallskip}
\end{tabular}
\par }
	\footnotesize{\textbf{Notes:}
	Column1: $r$-magnitude limit; the first \ps and \pc rows correspond to the full, not magnitude-limited samples.
	Column 2: OTELO-Deep image detection completeness from \citet{bon19}.
	Column 3: sample size.
	Column 4: sample baseline.
	Columns 5, 6 and 7: \ur, LDA and \dnn accuracies, respectively, obtained after 100 bootstrap runs.}
\end{table*}

We check our machine learning algorithms using flux-limited samples.
Table~\ref{tab:otelo_samples} shows the results of 100 bootstrap runs on different OTELO $r$-magnitude-limited samples.
We highlight the fact that all the \ur color, LDA, and \dnn accuracies are consistent within the errors.
However, for brighter samples, we can observe a downward trend in accuracies and upward trend in uncertainties for the LDA and \dnn classifications, while the \ur color results remain basically without change.
Analogously, Table~\ref{tab:cosmos_samples} shows the limit $r$ magnitude (Col. 1), the S\'{e}rsic index (Col. 2), the baseline (Col. 3), and the \dnn accuracy for the COSMOS \ps and \pc samples.
In this case, the training set size is always \ntrainC for the \ps  and \ntrainCc for the \pc samples, except for the \pc limit magnitude $r \leq 22$ with a sample size of 9852, for which the training set was set to \ntrainC.
As in the OTELO case, we notice the consistency of the \dnn accuracies within the errors, and the trend towards lower accuracies and larger uncertainties. 

There are two effects that may account for the trends in accuracy and uncertainty observed in the LDA and \dnn classification methods.
On one hand we detect a tendency for a lower proportion of LT galaxies in brighter galaxies, indicated by the baseline decrease.
As the \emph{a priori} probabilities of a galaxy to be ET or LT are more alike, the uncertainty in the classification increases. 
On the other hand, as the sample size shrinks in brighter samples, so do the fractions of the sample reserved for training and testing (70\% and 30\%, respectively).
This shrinking of the sample size leads to a less satisfactory training and a less precise testing.
Both effects, the baseline decrease and the sample shrinking, tend to reduce the classification accuracy.
With respect to the \ur classification, the \ur color discriminant is determined by low-redshift galaxies (see Figs.~\ref{fig:ur_z_sp} and \ref{fig:ur_z_cp}) that tend to dominate flux limited samples. 
Thus, the discriminant remains constrained around a value of 2, and the classification accuracy remains around 0.96. 
For the brighter \ps and \pc samples, with magnitude limit $r \leq 25$, the LDA and the \dnn accuracies are similar to that of the \ur color.
In the other two magnitude-limited cases ($r \leq 26$ and $r \leq 27$), the \dnn presents the highest accuracy, and the accuracy of the LDA is higher than the \ur color.


These results show that all the machine learning methods for classification presented in this paper are robust for both limited and unlimited flux samples.

\section{Conclusions}\label{sec:conclus}

Neural networks are becoming increasingly important for image classification and will play a fundamental role in mining future databases.
However, many of the current astronomical databases consist of catalogs of tabulated data.
Machine learning techniques are often used to analyze astronomical tabulated data, but analysis through \dnns is far less frequent and limited to low-redshift galaxies.

Here, we provide a consistent and homogeneous comparison of the popular techniques used in the literature for binary ET and LT morphological type classification of galaxies up to redshift $z \leq 2$.
We used data from the OTELO catalog for classifying galaxies by means of (i) the single \ur color discriminant, (ii) LDA using \ur color and the shape parameter (S\'{e}rsic or concentration index), and (iii) \dnn fed by visual-to-NIR photometry and shape parameter. 
We also applied the \dnn architecture developed for OTELO  on COSMOS to probe its reliability and reproducibility in a different database.

Both S\'{e}rsic index and concentration index shape parameters yield comparable results, but using the concentrations allowed to increase the size of OTELO and COSMOS available data.
All the machine learning methodologies for galaxy classification tested in this paper are robust and produce comparable results for both limited and unlimited flux samples.
Accuracy, \sensitivity, and \specificity estimates show that \dnn outperforms the other two methods and allows the user to classify more objects because of the missing data management.

These results show that \dnn classification is a powerful and reliable technique to mine existing optical astronomical databases.
For unresolved objects, the morphological identification is unattainable, the spectrum of a dim object is very difficult to obtain, and multiwavelength data are usually unavailable.
For most objects, photometric visible and near infrared observations are the only (and usually incomplete) accessible data.

This study indicates that \dnn classification may address the mining of currently available astronomical databases better than other popular techniques.

\begin{table}[!tp]
\caption{Mean accuracy for COSMOS samples at different magnitude limits in the r-band} 
\label{tab:cosmos_samples}
\centering
\begin{tabular}{lrr@{$\,\pm\,$}lr@{$\,\pm\,$}l}
  \noalign{\smallskip}
  \hline
  \noalign{\smallskip}
  $r_{lim}$ & $n$ & \multicolumn{2}{c}{ Baseline } & \multicolumn{2}{c}{ DNN } \\
  \noalign{\smallskip}
 \hline
 \noalign{\smallskip}
  \multicolumn{6}{c}{COSMOS SP sample} \\
  \noalign{\smallskip}
  25.80$^1$ & \numprint{34688} & 0.835 & 0.002 & 0.972 & 0.003 \\ 
  24.00     & \numprint{34128} & 0.836 & 0.002 & 0.969 & 0.003 \\ 
  23.00     & \numprint{23562} & 0.841 & 0.002 & 0.971 & 0.003 \\ 
  22.00     & \numprint{ 9368} & 0.783 & 0.004 & 0.966 & 0.004 \\ 
  \noalign{\smallskip}
  \multicolumn{6}{c}{COSMOS CP sample} \\
  \noalign{\smallskip}
  26.88$^1$ & \numprint{105758} & 0.906 & 0.001 & 0.971 & 0.002 \\ 
  24.00     & \numprint{ 65808} & 0.896 & 0.001 & 0.976 & 0.001 \\ 
  23.00     & \numprint{ 25912} & 0.850 & 0.002 & 0.972 & 0.003 \\ 
  22.00     & \numprint{  9852} & 0.787 & 0.004 & 0.964 & 0.004 \\ 
  \hline
    \noalign{\smallskip}
  \multicolumn{6}{l}{\footnotesize{$^1$ Sample not limited in magnitude.}}
\end{tabular}
\end{table}

An important limitation for all machine learning techniques is the availability of labeled data, that is, data that have already been classified or measured.
This limited us to a binary ET and LT classification and to impose a redshift threshold.
Incorporating reliable synthetic data for classification training is an important goal if we wish to overcome these limitations.

Our results provide compelling support for extending the \dnn classification to targets other than binary morphological classification of galaxies, such as separating stars from galaxies, deciphering the spectral type of stars, and detecting  rare events.
The application of \dnn is not restricted to classification problems.
Our results strongly suggest that \dnn methods can also be very effective in exploring other issues such as, for example, photometric redshift estimates.

\begin{acknowledgements}
The authors are grateful to the referee for careful reading of the paper and valuable suggestions and comments.
This work was supported by the project Evolution of Galaxies, of reference AYA2014-58861-C3-1-P and AYA2017-88007-C3-1-P, within the "Programa estatal de fomento de la investigaci\'{o}n cient\'{\i}fica y t\'{e}cnica de excelencia del Plan Estatal de Investigaci\'{o}n Cient\'{\i}fica y T\'{e}cnica y de Innovaci\'{o}n (2013-2016)" of the "Agencia Estatal de Investigaci\'{o}n del Ministerio de Ciencia, Innovaci\'{o}n y Universidades", and co-financed by the FEDER "Fondo Europeo de Desarrollo Regional".\\
JAD is grateful for the support from the UNAM-DGAPA-PASPA 2019 program, the UNAM-CIC, the Canary Islands CIE: Tricontinental Atlantic Campus 2017, and the kind hospitality of the IAC. \\
MP acknowledges financial supports from the Ethiopian Space Science and Technology Institute (ESSTI) under the Ethiopian Ministry of Innovation and Technology (MoIT), and from the Spanish Ministry of Economy and Competitiveness (MINECO) through projects AYA2013-42227-P and AYA2016-76682C3-1-P.\\
APG, MSP and RPM were supported by the PNAYA project: AYA2017--88007--C3--2--P.\\
MC and APG are also funded by Spanish State Research Agency grant MDM-2017-0737 (Unidad de Excelencia Mar\'{\i}a de Maeztu CAB).\\
EJA acknowledges support from the Spanish Government Ministerio de Ciencia, Innovaci\'{o}n y Universidades though grant PGC2018-095049-B-C21. \\
M.P. and E.J.A also acknowledge support from the State Agency for Research of the Spanish MCIU through the ”Center of Excellence Severo Ochoa” award for the Instituto de Astrof\'{\i}sica de Andaluc\'{\i}a (SEV-2017-0709).\\
JG receives support through the project AyA2018-RTI-096188-B-100.\\
MALL acknowledges support from the Carlsberg Foundation via a Semper Ardens grant (CF15-0384).\\
JIGS receives support through the Proyecto Puente 52.JU25.64661 (2018) funded by Sodercan S.A. and the Universidad de Cantabria, and PGC2018--099705--B--100 funded by the Ministerio de Ciencia, Innovaci\'{o}n y Universidades.\\
Based on observations made with the Gran Telescopio Canarias (GTC), installed in the Spanish Observatorio del Roque de los Muchachos of the Instituto de Astrof\'{\i}sica de Canarias, in the island of La Palma. 
This work is (partly) based on data obtained with the instrument OSIRIS, built by a Consortium led by the Instituto de Astrof\'{\i}sica de Canarias in collaboration with the Instituto de Astronom\'{\i}a of the Universidad Aut\'{o}noma de M\'{e}xico. 
OSIRIS was funded by GRANTECAN and the National Plan of Astronomy and Astrophysics of the Spanish Government
\end{acknowledgements}




\end{document}